\documentclass[twocolumn]{aastex631}

\usepackage{natbib}
\usepackage{graphicx}
\usepackage{comment}
\usepackage{longtable}
\usepackage{amsmath,amsthm,amssymb,amsfonts}
\usepackage[caption=false]{subfig}

\newcommand{\ndsfg}{71}
\newcommand{\fluxlimit}{8~mJy}
\newcommand{\herschel}{\textit{Herschel}}
\newcommand{\planck}{\textit{Planck}}

\defcitealias{Gralla20}{G20}
\defcitealias{Reuter20}{R20}
\defcitealias{Swinbank14}{S14}
\defcitealias{Magnelli12}{M12}

\begin{document}
\title{A flux-limited sample of dusty star-forming galaxies from the Atacama Cosmology Telescope: physical properties and the case for multiplicity}

\author[0000-0002-4176-845X]{Kirsten R. Hall}
\affiliation{Radio and Geoastronomy Division, Center for Astrophysics $\vert$ Harvard \& Smithsonian, Cambridge, MA 02135, USA \\} 
\author[0009-0004-5685-6155]{Jake B. Hassan}
\affiliation{Department of Physics and Astronomy, Stony Brook University, Stony Brook, NY 11794, USA \\} 
\author[0000-0002-9330-8738]{Richard M. Feder}
\affiliation{Berkeley Center for Cosmological Physics, University of California, Berkeley, CA 94720,
USA}
\affiliation{Lawrence Berkeley National Laboratory, Berkeley, California 94720, USA}
\author[0000-0003-4496-6520]{Tobias A. Marriage}
\affiliation{The William H. Miller III Department of Physics and Astronomy, Johns Hopkins University, 3701 San Martin Drive, Baltimore, MD 21218, USA \\} 
\author[0000-0001-8253-1451]{Michael Zemcov}
\affiliation{School of Physics and Astronomy, Rochester Institute of Technology, Rochester, NY 14623 \\} 
\affiliation{Jet Propulsion Laboratory, 4800 Oak Grove Drive, Pasadena, CA 91109, USA \\}
\author[0000-0003-3191-5193]{Jesus Rivera}
\affiliation{Department of Physics and Astronomy, Swarthmore College, Swarthmore, PA 19081, USA \\} 

\begin{abstract}
We report the modeling of the millimeter and far-infrared spectral energy distributions of \ndsfg{} dusty star-forming galaxies (DSFGs) selected by the Atacama Cosmology Telescope (ACT) with a 5$\sigma$ flux-density limit of \fluxlimit{} at 220~GHz (1.4~mm). 
All sources were cross-identified with \herschel{} surveys at 500, 350, and 250 $\mu$m, and nineteen of our sources were observed with the Submillimeter Array. 
A probabilistic cataloging algorithm, PCAT, favors multiple unresolved flux components in the \herschel{} data for the majority of ACT-selected DSFGs. 
We compare the derived physical properties of the DSFGs obtained from modeling the flux densities with those from similar studies of DSFG populations.
We find the median, 16th and 84th percentiles for the following SED-model parameters: redshift $z_\mathrm{phot}=3.3^{+0.7}_{-0.6}$, apparent size $\sqrt{\mu} d=5.2^{+0.9}_{-2.4} \; \mathrm{kpc}$, apparent dust mass $\log_{10} (\mu M_d/M_\odot)=9.14^{+0.12}_{-0.04}$ and cutoff temperature $T_c=35.6^{+4.8}_{-1.6} \; \mathrm{K}$, for a power law distribution, and the corresponding apparent Far-IR luminosity $\log_{10}(\mu L_\mathrm{IR}/L_\odot)=13.6^{+0.2}_{-0.3}$, where $\mu$ is lensing magnification. 
While many of the properties broadly agr ee with those of samples of primarily lensed DSFGs, we exercise caution in interpreting them. ACT's lower flux limit, the PCAT decomposition, and the SMA observations all suggest that some fraction of these DSFGs are likely to be unlensed and possibly multiples. The SMA data indicate that at least fourteen out of nineteen sources are such, either via ``missing" flux in comparison to the model or detection of additional sources in the fields. Additional high-resolution follow-up and redshift determination are needed to better understand this flux-limited sample of DSFGs.
\end{abstract}

\section{Introduction}

Infrared-luminous, dust-enshrouded galaxies, or more simply, dusty star forming galaxies (DSFGs) are increasingly understood to be the primary contributors to the cosmic star formation rate density (at least out to $z\sim4$) and sites of protocluster environments \citep{char01,grup13,clem14,plan15a, mack17,hall18, oteo18,gree18, gome19,Grup20,zava21}.
While measurements of the rate and nature of star formation of cosmological volumes extend as far back as 13.7 billion years ($z\approx11$), they are dominated by rest-frame ultraviolet light \citep{Schenker13, Bowens15, oesc18} and face the challenge that a significant fraction of starlight may be absorbed by dust in the host galaxy. 
Therefore, understanding the redshift distribution and overall nature of dust-dominated objects is essential for a complete understanding of the formation and evolution of galaxies over cosmic time. 

Observations at millimeter wavelengths have the advantage of the negative K-correction, making galaxies of fixed luminosity and dust temperature have approximately constant observed millimeter flux densities between $z=1-7$ \citep{blai02,Casey14}. 
Wide-field surveys from far-infrared and millimeter-wavelength single-dish or single-aperture observatories, such as the \herschel{} Space Observatory\footnote{\herschel{} is an European Space Agency space observatory with science instruments provided by European-led Principal Investigator consortia and with important participation from NASA.}, the Atacama Cosmology Telescope (ACT), the South Pole Telescope (SPT), and the \planck~satellite, are therefore crucial for studying the dusty galaxies responsible for the total cosmic infrared background emission. 
Targeted observations and smaller surveys from the instruments Submillimetre Common-User Bolometer Array (SCUBA, and SCUBA-2) on the James Clerk Maxwell Telescope have also proven exceedingly fruitful in studying DSFGs.
However, until recently, detailed source characterization has been limited by confusion due to the instruments' low spatial resolution. 

Results from the 2500~deg$^2$ survey from SPT have unveiled $\sim$500 dust-dominated galaxies \citep{ever20}, 81 of which were identified as strongly lensed systems and have been spectroscopically followed up at high resolution using ALMA \citep{Reuter20}. 
These sources lie at the highest extreme of the infrared luminosity distribution of all known DSFGs, selected with a 1.4~mm flux density limit of 20~mJy. 
The median redshift of the sample is $z=3.9$, exemplifying the power of millimeter-wave observations for identifying and characterizing $z\gtrsim$~3 DSFGs \citep[see also,][]{gree19,yan20,yan22,quir24}. 


Higher resolution ($\sim$1") surveys \citep{dunl17,case21} and follow-up observations \citep{Karim13, Swinbank14, daCunha15} of extremely red and infrared luminous sources detected in millimeter and far-infrared surveys are revealing that these sources are either gravitationally lensed, amplifying their apparent far-infrared luminosities (as in \citealt{Reuter20} or many of the sources in \citealt{garr23}) or they are multiple, blended galaxies in the lower-resolution beams \citep[e.g.,][]{wang21,calv23,garr23}.
Such surveys with interferometers are crucial for high spatial resolution and spectroscopic characterization of this key population of galaxies, but the surveys are time consuming and the time available is limited.
Thus, there exists a nice synergy with catalogs of hundreds of sources extracted from millimeter-wavelength, low-resolution, single-aperture/dish surveys.

Recently, \cite{quir24} cross-matched red-\herschel{} sources with publicly available ALMA data to create a sample of $>$2400 DSFGs observed at 1.3~mm and high spatial resolution ($\sim$1"). 
They find that 20\% are multiple galaxy systems, and 5\% are either lensed or close ($\leq$3") mergers.
Similar percentages of multiples have been reported by studies of smaller samples \citep{Ma19,gree20,mont21,bend23,cox23,cair23,clem24}.
Some fraction of the multiple detections may be associated galaxies in over-dense environments at high-redshift, or protoclusters, such as the systems found in \citet{oteo18} and \citet{jone24}.

Continuing to expand the characterization of the DSFG populations at lower flux densities and assess their lensed or multiplicity potential is the aim of this work. 
Of the 644 extragalactic sources in the latest ACT source catalog extracted from the equatorial survey \citep{Gralla20}, 268 are classified as DSFGs (distinguished from AGN, or strong synchrotron sources, although there will be overlap in these populations). 
The ACT equatorial survey reached standard errors of 2-3~mJy in the 218~GHz band.
At this depth, many of the DSFGs are likely to be fainter, unlensed systems enabling us to probe the physical properties of a portion of the DSFG population that is not at the extreme tail of the luminosity function \citep{Negrello10, quir24}. 
Moreover, the equatorial region of ACT has significant overlap with the two largest \herschel{} survey fields, allowing far-infrared through millimeter characterization of the cold dust emission in these galaxies.

In this work, we model the spectral energy distributions (SEDs) of 71 ACT-selected DSFGs that also have far-infrared photometry from \herschel{}. 
Using the locations of the ACT-selected sources as a rough prior, we perform blind source detection and photometry using the forward modeling tool ``Probabilistic Cataloging" \citep[PCAT,][]{Brewer2013,port17,Feder20}. 
We then use the extracted \herschel{} flux densities of putative counterparts in combination with ACT to model the broad-band SEDs of the sources, inferring the photometric redshift, cutoff dust temperature, apparent dust mass, size, and far-infrared (FIR) luminosity, and we report on the number of 
flux components predicted by PCAT.

Further, we provide flux densities and high-resolution imaging for a subset of 19 of our sources using observations from the Submillimeter Array. These flux densities are compared with what is expected from the model SED, and we make an assessment on the possibility for multiplicity. 
This study is part of a series on millimeter-wave sources discovered by ACT \citep{Marriage11,Marsden14,Gralla20,vargas23}. 
In particular, this paper expands and augments our work in \cite{Su17}, in which the combined ACT-\herschel{} SEDs of nine DSFGs were analyzed.

In Section~\ref{sec:data} we describe the multi-wavelength and multi-facility data set for our sample, and in Section~\ref{sec:analysis} we detail our analysis including derivation and modeling of the SEDs. We report our modeling results and discuss the findings including a comparison with similar studies and the assessment of the SMA flux densities and PCAT-predicted multiplicity in Section~\ref{sec:results}. Then, we summarize our findings and conclusions in Section~\ref{sec:conclusion}. 

\section{Data}
\label{sec:data}

\begin{figure*}[ht!]
\centering
\includegraphics[width=\textwidth]{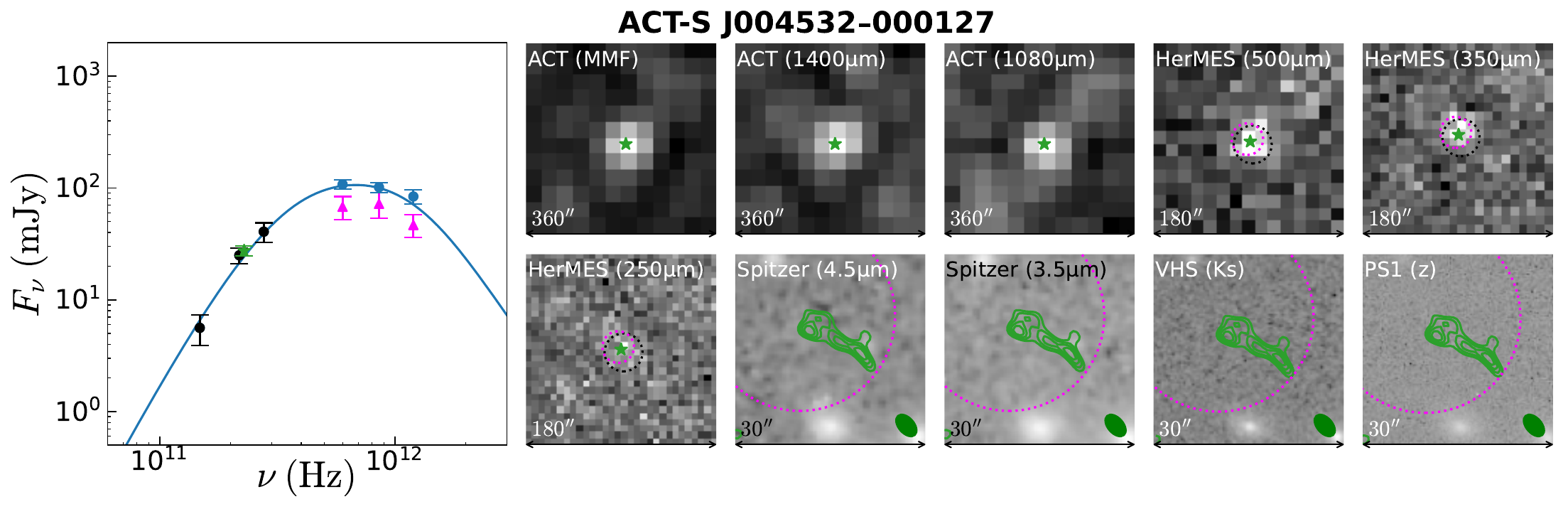}
\caption{The SED of ACT-S J004532-000127  with imaging at different wavelengths (right). Left: Black points are the ACT de-boosted fluxes, blue points are the \herschel{} ensemble fluxes, magenta triangles are the PCAT brightest flux component, and the green star is the summed flux from the SMA detections. The blue curve is the best-fitting SED derived from the medians of the posterior distributions of the parameters. Right: Imaging data from the microwave to optical centered on the ACT source location. Green stars on the ACT and \emph{Herschel} thumbnails indicate the location of the SMA source closest to the ACT source location, and we plot the 3, 5 and 7$\sigma$ ($\sigma$=0.8~mJy) SMA contours in green on the \emph{Spitzer}, VHS, and PS1 images. The green ellipse at the bottom right of the high-resolution data shows the shape of the synthesized SMA beam, which for these data is 2.5"x4.1". The same figure for the other sources that were observed by the SMA are placed in a Figure set (19 images) and available in the online journal.}
\label{fig:thumbnail}
\end{figure*}

Our DSFG sample is a subset of the catalog from \citet{Gralla20} (hereafter \citetalias{Gralla20}), which was extracted from ACT's 480~deg$^{2}$ sky survey over the celestial equator. This region was chosen to overlap with the Sloan Digital Sky Survey's deep imaging Stripe 82 \citep[e.g.,][]{Jiang2014}, which benefits from a broad range of complementary observations.
In particular, the ACT data used in \citetalias{Gralla20} have 120~deg$^{2}$ of overlap with the \herschel{} sky surveys: the \herschel{} Stripe 82 Survey (HerS, \citealt{Viero14}) and the HerMES Large Mode Survey (HeLMS, \citealt{Oliver12}).


\subsection{The Atacama Cosmology Telescope (ACT)}

The Atacama Cosmology Telescope was a 6-meter diameter telescope observing from 5,190~m elevation in the Atacama Desert, Chile \citep{fowler07}. 
ACT first observed over 1000~deg$^2$ of the southern and equatorial sky at 148~GHz (2000~$\mu$m), 218~GHz (1400~$\mu$m), and 277~GHz (1080~$\mu$m) with the Millimeter Bolometric Array Camera (MBAC) from 2007-2010 \citep{Swetz11}.
The angular resolution of the telescope at these bands was 1.4\arcmin, 1.1\arcmin, and 0.9\arcmin, respectively, and the beam, data reduction, and map calibration details can be found in \citet{dunner13} and \citet{hasselfield13}. 
ACT receivers were then upgraded twice for two successive generations of observations for the ACTPol \citep{Thornton16} and Advanced ACTPol \citep{Henderson15} experiments, but the catalog used here from \citetalias{Gralla20} is entirely drawn from the original ACT MBAC-based survey.\footnote{The ACTPol frequency bands were 90 and 150~GHz, limiting that survey's contribution to DSFG studies. The 16,000 deg$^2$ Advanced ACTPol survey at 220~GHz will yield an interesting sample of bright DSFGs once the data is reduced to a catalog. Therefore, the catalogs from \citetalias{Gralla20} and \cite{vargas23} remain the primary ACT DSFG assay, with the \citetalias{Gralla20} sample having the advantage of multi-frequency coverage in Stripe 82.}

We refer the reader to \citetalias{Gralla20} for the complete description of the survey area and DSFG source extraction. 
Of the 98 ACT DSFG candidates with \herschel{} coverage, 24 sources were identified as nearby galaxies. \citetalias{Gralla20} applied the ``nearby'' classification if a galaxy is visually well-resolved in optical imaging data.
Another three putative DSFGs appear to be of Galactic origin or contain Galactic contamination (ACT-S J010838-014328, ACT-S J010501+020847, and ACT-S J021604-004026). 
Excluding these 27 sources gives 71 ACT-selected sources for this study (42 in HerS, 30 in HeLMS, and one which is in both). 
Of these \ndsfg{} DSFGs, 63 were selected using a multi-frequency matched filter (MMF) that assumed a thermal-dust like spectrum. 
The remaining 8 sources were selected based on the match-filtered 218~GHz dataset alone.
Sources with signal-to-noise ratio SNR $>5$ are included in the catalog. 
Because the noise is relatively uniform in the ACT maps, this results in an approximately flux-limited sample with 5$\sigma$ limiting raw (not de-boosted) 218-GHz flux densities of 9.5~mJy for the MMF selection and 12~mJy for the 218~GHz-only selection.
The ACT flux densities and uncertainties used in this study and provided in Table \ref{tab:catalog} are the de-boosted quantities \citep{Gralla20b} provided in \citetalias{Gralla20} with an effective 5$\sigma$ flux limit of 8~mJy.
At this survey depth, the sample is expected to comprise a mix of lensed and unlensed systems at $z>1$ \citep[e.g.,][]{Negrello10, Bethermin11,quir24}.

\subsection{\herschel{} Space Observatory}
\label{HSO}

Cross-matching our sources with far-infrared data from the SPIRE Instrument of the \herschel{} Space Observatory \citep{pilb10} is particularly useful for measuring the peak of the thermal dust emission spectrum of the DSFGs. 
The \herschel{} mission was equipped with three far-infrared observing instruments: the Photodetector Array Camera and Spectrometer (PACS, \citealt{pogl10}), the Spectral and Photometric Imaging Receiver (SPIRE, \citealt{grif10}), and the Heterodyne Instrument for the Far Infrared (HIFI, \citealt{graa10}). 
PACS and SPIRE were also equipped with cameras, and the observations span 2009 to 2013. 
We exclusively use SPIRE data here due to its observational overlap with the ACT equatorial field.

\herschel{}'s SPIRE instrument observed at 250~$\mu$m (1200~GHz), 350~$\mu$m (857~GHz), and 500~$\mu$m (600~GHz) \citep{griffin10}. 
HerS \citep{vier14} is a 79 ~deg$^{2}$ survey over the celestial equator ($13^{\circ} < \alpha < 37^{\circ}, -2^{\circ} < \delta < 2^{\circ}$), and HeLMS \citep{Asboth16} is a larger survey (approximately 280~deg$^2$), overlapping the equator and extending further in declination (roughly $-10^{\circ} < \alpha < 18^{\circ}, -8^{\circ} < \delta < 8^{\circ}$). 
Details of the SPIRE beam are found in \citet{griffin13}; the full widths at half maxima (FWHM) are 18.1$''$, 25.2$''$, and 36.6$''$ for the 250~$\mu$m, 350~$\mu$m, and 500~$\mu$m bands, respectively. 
\citet{Viero14} reported the total flux density uncertainty, including instrumental and confusion noise, for HERS as $14.8$, $12.9$, and $13.0$~mJy/beam for the 500, 350, and 250~$\mu$m bands, respectively. For the total noise in HELMS, \cite{Asboth16} reported $10.45$, $12.88$, and $15.6$~mJy/beam in the 500, 350, and 250~$\mu$m bands, respectively. 
We further discuss the relevant noise values in the context of flux extraction in Section~\ref{sec:SPIRE}. 


The maps used for this work have been generated using the final version of the \herschel{} Interactive Processing Environment (HIPE; \citealt{HIPE2011}) and its associated v14.3 calibration tree.  
With this processing, the absolute calibration in each band has been estimated to have a 1.5\% statistical uncertainty, with an additional 4\% uncertainty arising from the model for Neptune \citep{Bendo2013}. The processed time streams are transformed into maps using the SHIM map making algorithm \citep{Levenson2010}, which has been shown to have unity transfer function on the scales of interest in this program \cite{Viero2013}. Additional details about the data processing are available in \citet{zemc24} and the maps themselves are publicly available for use on Zenodo\footnote{Persistent DOI: 10.5281/zenodo.13352296}.

\subsection{Submillimeter Array (SMA)}
\label{sec:sma}

The Submillimeter Array is an interferometer made up of 8 6-m dishes located at 4100-m at the summit of Maunakea, Hawai'i.
The array has two receivers, each currently having a total bandwidth of 24~GHz, that can be tuned to separate frequencies or the same frequency for maximum signal-to-noise. 
Prior to 2019, the receivers had a bandwidth of 16~GHz.
All of the observations used in this work were obtained with dual tuning.

Nineteen of the DSFGs in this catalog were observed with the Submillimeter Array during two different observing cycles in 2013 (Project ID: 2013A-S005, P.I. Clements; Project ID: 2013B-S066, P.I. Baker), a third cycle in 2017 (Project ID: 2017A-S042, P.I. Rivera), and a fourth in 2023 (Project ID: 2023A-S049, P.I. Hall).
The 2013 and 2017 observations were obtained before the receiver upgrade and thus have a total bandwidth of 16~GHz, and the continuum observations have effective frequencies of $\nu_{eff}=270$~GHz (2013A), 228.5~GHz (2013B) and 225~GHz (2017) and RMS sensitivities of $\sigma_{rms} \approx$~1.8~mJy (2013A), 1.2~mJy (2013B), and 0.49~mJy (2017).
These data were calibrated using the MIR software package in IDL, then converted to the Common Astronomy Software Applications (CASA, \citealt{mcmu07}) package measurement sets for imaging and analysis. 
Each of these 2013 and 2017 observations have approximately 4" resolution as they were observed in the SMA's second most compact configuration, ``COM". 
Ten of the sources were observed in the 2023A cycle at $\nu_{eff}=225$~GHz and RMS sensitivities of $\sigma_{rms} \approx$~0.8 mJy. 
The raw observations were converted to CASA measurement sets using the publicly available Python package pyuvdata \citep{hazel2017}. 
Calibrations and imaging were then performed in CASA. 
These data were observed in the SMA's most compact configuration, ``SUB", with varying numbers of antennas and have spatial resolution ranging from $\sim$6-9". 

\section{Analysis}
\label{sec:analysis}
We derived 148-1200~GHz SEDs using ACT and \herschel{} observations. 
The \herschel{}-SPIRE flux densities are the sum over an ensemble of unresolved components within the ACT beam as determined by PCAT.
We modeled the ensemble SEDs with modified black body (MBB) spectra, and compared the single aperture derived ACT flux densities with those from high-resolution interferometric observations with the SMA. 
We use the comparison to assess whether the PCAT-derived components and flux densities correspond to multiple source counterparts.

In Figure~\ref{fig:thumbnail}, we show an example of all of the data for a one of our sources,
ACT-S J004532-000127. 
For each source, we derived the SED from the ACT and \herschel{} cutouts, and we include in the plot the flux density of the brightest PCAT component (magenta) and the flux densities for the nineteen sources observed with the SMA (green).
We also include the near-IR and optical cutouts, as available, for possible detections or foreground lensing galaxies, and examine where the SMA location and contours fall on the low- and high-resolution data, respectively.
Analogous figures for the other eighteen SMA-observed sources are shown in the online journal in Figure set 1. 
If the source has a VLA detection, its location is indicated with a red pentagon.
Note that the majority of the optical and near-IR thumbnails are 30" across, but in a few exceptions in which the SMA source is more distant, the size scale increases to 40" or in one case, 60".

\subsection{Deriving SEDs}
\label{sec:SEDs}

We used the de-biased ACT flux densities from \citetalias{Gralla20} for our primary analysis. These are corrected for Eddington bias and other selection effects \citep{Gralla20b}.
We only used the ensemble \herschel{} flux densities in the SED modeling.
The ensemble SED corresponds to the hypothesis that the source is a single source (possibly lensed).
It is also useful to the extent that fitting the total flux density from a group of DSFGs with similar properties, such as dust temperature, using a single SED model is valid. This assessment breaks down, however, if the ensemble flux derives from sources at different redshifts or with different properties.

Figure~\ref{fig:combination} shows the SEDs for all 71 of our sources along with the fits to the ensemble fluxes (see Section~\ref{sec:modeling} for our model definition). 
Five of the sources have spectroscopic redshifts via additional follow-up observations: J0022–0155 \citep[$z=5.161$]{Asboth16}, J0044+0118 \citep[$z=4.163$]{Ma19}, J0107+0001 \citep[$z=3.3327$]{Rivera19}, J0116–0004 \citep[$z=3.7908$]{Rivera19}, and J0209+0015 \citep[$z=2.5534$]{Geach15,Su17}. 
We performed an additional fit to these five sources with the redshift fixed to its spectroscopic value.
Nineteen of the sources have flux densities extracted from the SMA images as described below (Section~\ref{sec:smaflux}) and are indicated as green stars on the SEDs.


\subsubsection{SPIRE source detection and flux density estimation}
\label{sec:SPIRE}

We extracted $5'\times5'$ cutouts around each of the ACT source locations and identified \herschel{}-SPIRE source components within a $30\arcsec$ radius from each of our ACT-selected DSFGs.\footnote{We use the neutral word ``component'' for unresolved (point) sources identified by the PCAT algorithm because these may be associated with one \herschel{} counterpart with a flux distribution that does not match a \herschel{} point spread function or multiple distinct counterparts.} This radius was chosen to be three times the approximate astrometric uncertainty for the fainter sources in the ACT catalog \citepalias{Gralla20}. 
We used the tool \texttt{PCAT-DE}\footnote{PCAT-DE is publicly available on GitHub at https://github.
com/RichardFeder/pcat-de, with corresponding documenta-
tion} (PCAT) to perform point source detection and de-blending.
PCAT is accomplished through Bayesian image forward modeling methods \citep{Feder20,Feder23}. 
The algorithm employs trans-dimensional sampling to estimate a posterior source flux density distribution that is marginalized over blending degeneracies and low-SNR neighbors, and can return flux posteriors for individual sources. 
\citet{Feder23} tested the effectiveness of PCAT on synthetic Herschel-SPIRE multi-band observations. They successfully perform point-source detection in the presence of diffuse background emission and recover infrared galaxy counts cases of significant contamination. 

Here we use PCAT on the \herschel{}-SPIRE maps by jointly modeling the three-band SPIRE cutouts corresponding to each ACT source, and using the derived WCS solutions to forward model point sources across maps. 
This enables us to more effectively detect and model the contributions of faint-end sub-mm sources, and to estimate source colors without the need to cross-match single-band catalogs, which can introduce additional errors. 

Across the 71 ACT-selected regions, the SPIRE maps have median instrument noise levels of 13, 12, and 15~mJy~beam$^{-1}$ for 250~$\mu$m, 350~$\mu$m and 500~$\mu$m, respectively, with a larger dispersion across cutouts in the HeRS footprint, however this is captured by our noise model across each individual cutout. 
Confusion noise in the SPIRE maps is estimated to be between $5-6$ mJy beam$^{-1}$ at $S_{min}=30$ mJy \citep[see Fig. 5 of][]{Nguyen10}, i.e., the maps we analyze are largely instrument noise-dominated. 
We imposed a minimum flux density of $S_{min}^{500} = 30$ mJy.
This is lower than employed by traditional source detection methods; however, within the PCAT framework it enables more effective modeling of sub-threshold confusion noise. 
Furthermore, due to the simultaneous multi-band fitting capabilities of \texttt{PCAT-DE}, the $2\sigma$ threshold is a lower bound to the effective detection SNR. 
For 250~$\mu$m and 350~$\mu$m we only imposed a flux density positivity prior, and we adopted an uninformative color prior for the fits as the goal is to measure the thermal blackbody peak traced by the \herschel{} bands, the position of which is not known \emph{a priori}. 


We performed Metropolis-Hastings sampling with \texttt{PCAT-DE} on each set of cutouts for $10^6$ iterations, after which the chains are thinned by a factor of 1000 to reduce the autocorrelation length of samples. 
We used the last 500 thinned samples from each chain, although we observed that the chains typically converge after the first 100-200 thinned samples. 
In general, our model is a good fit to the SPIRE data -- the median reconstruction reduced $\chi^2$ statistic, $\chi^2_{red}$, for the observed maps is 1.09, 1.06 and 1.10 for 500 $\mu$m, 350 $\mu$m and 250 $\mu$m, respectively, with only $\sim 6$ ACT-selected postage stamps with $\chi^2_{red} > 1.5$. 
The median $\chi^2$ across the cutouts shows no trends with either the number of \herschel{} flux components per ACT source, as defined earlier, nor the total number of PCAT-detected components in each set of $5'\times5'$ cutouts. 

The PCAT posterior is represented as an ensemble of source catalog realizations. 
For each ACT source, we define two flux density quantities to characterize the \herschel{} fluxes from the PCAT output. 
The first is an ``ensemble" flux density -- for each thinned sample, we compute the summed flux density for all flux components within 30\arcsec\ of the ACT source position, and repeat for all samples to collect a flux posterior.
The second is a ``brightest source" flux density, which follows a similar procedure but for which only the brightest model source at 500 $\mu$m is selected at each sample. 
The brightest flux is shown only for comparison as magenta triangles in the SEDs of Figures~\ref{fig:thumbnail} and \ref{fig:combination}, and not included in the SED fitting analysis. 

\subsubsection{Consistency tests for SPIRE photometry}
We tested the sensitivity of the SPIRE photometry by varying parameters of the \texttt{PCAT-DE} run configuration, after which we compared the recovered ensemble and brightest source flux density estimates. 
In particular, we tested:
\begin{itemize}
\item Perturbing $S_{min}$: For our fiducial selection band (500 $\mu$m), we varied $S_{min}=30$ by $\pm 5$ mJy. 
\item Varying the PCAT selection band: we performed a separate run in which we use 250 $\mu$m as the pivot flux density with $S_{min}=30$ mJy. 
\end{itemize}
For all tests, we found a close one-to-one correspondence of recovered fluxes as a function of $S_{\lambda}$. 
In general, we found that the ensemble flux densities are more stable to analysis variations than the brightest flux densities, though both have departures that are within the recovered 1$\sigma$ uncertainties. 
As expected, there is a small fraction of sources for which the different $S_{min}$ boundaries affect the recovered colors; however, these are largely low-SNR sources with large fractional uncertainties in all test configurations.  



One objective of this paper is to evaluate and interpret cases where there may be multiple \herschel{} flux components per ACT-selected DSFG. 
In addition to flux densities, we derive the ``multiplicity'' for each ACT source as the median number of \herschel{} flux components from the distribution of posterior samples within a 30" radius. Thus defined, we find that 13, 35, 18, 4, and 1 ACT sources have PCAT-determined components of 1, 2, 3, 4, and 6, respectively.
We discuss the viability (or lack thereof) and interpretation of these  multiplicities in Section~\ref{sec:multi}.

\subsection{SMA flux density estimation}
\label{sec:smaflux}

SMA counterparts are identified in the field of view via two means: (1) visual inspection of the dirty maps to identify sources, followed by cleaning to the threshold RMS of the map and (2) using auto-masking within CASA's ``tclean" function with varying SNR levels, going as low as 3.5$\sigma$ to search for potential low SNR counterparts in the dirty maps. In this case we also inspect the maps by eye to assess possible false detections due to aliasing of the PSF. We only consider as confident detections sources with SNR$\geq$4.

The source flux densities from the SMA are extracted by several means. 
If the source can be well approximated by a two-dimensional Gaussian, we use the CASA routine ``imfit" to extract the flux. 
This routine enables source extraction both for unresolved point sources and resolved sources that are well approximated by a two-dimensional Gaussian via the peak flux density and the integrated flux density, respectively. 
For sources that are not well defined by a two-dimensional Gaussian, for example, J001133-001835 and J011640-000457, we defined the source flux by summing over the pixels around the source central location contained within the 3$\sigma$ contour line and converted to a flux density (in mJy) using the beam size in pixels.

\begin{figure*}[ht!]
\epsscale{1.2}
\plotone{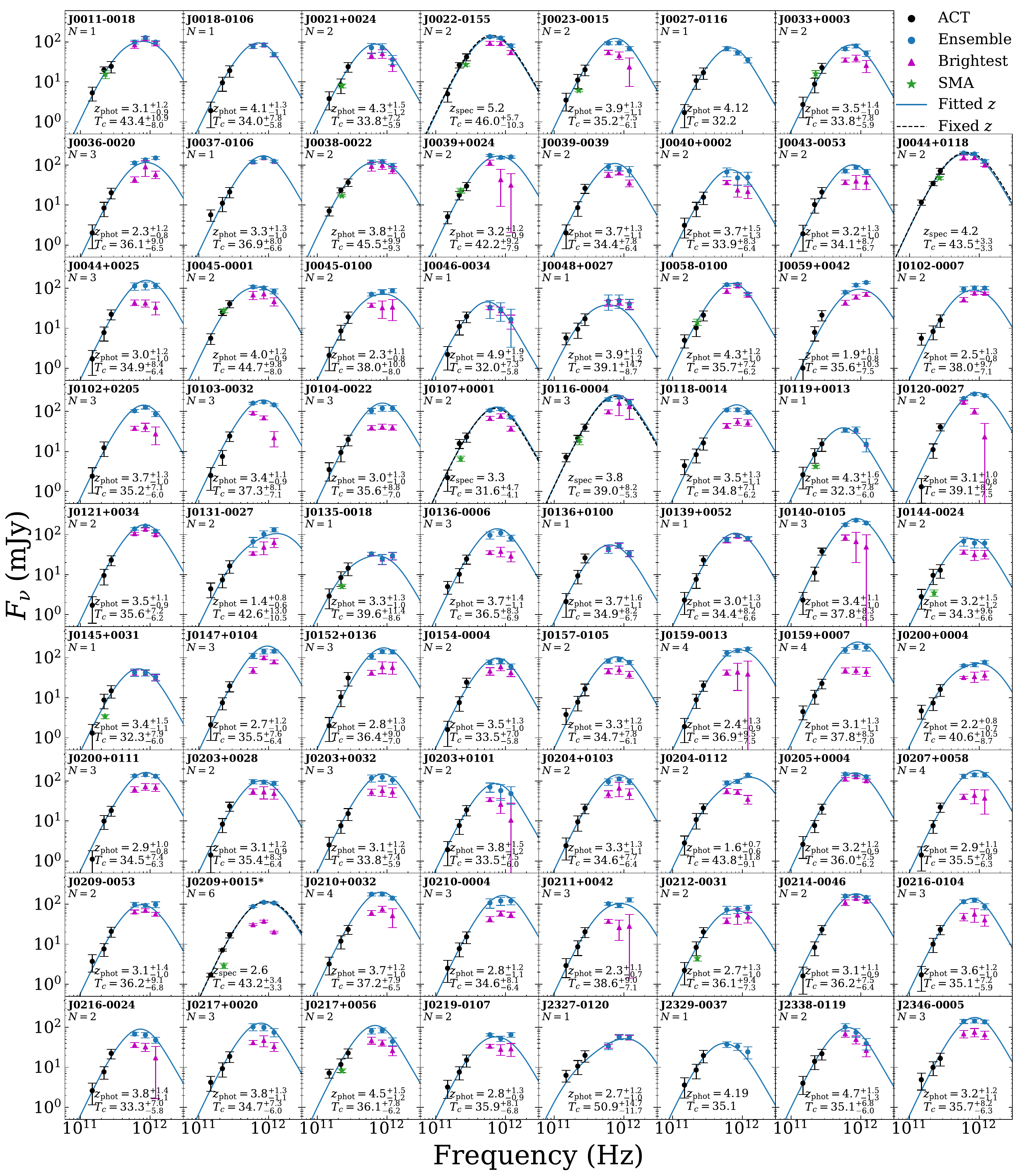}
\caption{SEDs for all 71 DSFGs in our sample. The ACT flux densities are shown as black points, the ensemble \herschel{} flux densities are blue points, and the brightest PCAT component flux densities are plotted as magenta triangles. 
For many sources, the brightest flux density is equal to the ensemble flux within uncertainty, indicating that the PCAT posteriors indicate only one point source flux component in the \herschel{} data. 
SMA flux densities are plotted as green stars. 
The blue curves are the best-fitting SEDs, including photometric redshift as a fit parameter, and for our five sources with spectroscopic redshifts, we plot the best-fitting fixed-z SED as a black dashed line. 
The median photometric redshift (or the spectroscopic redshift) and the median cutoff temperature are reported at the bottom of each figure, and the number of PCAT flux components within 30" is reported on the top left below the source name.
Uniquely, the fluxes for J0209+0015 are all divided by 10. 
}
\label{fig:combination}
\end{figure*}

The SMA location is indicated on the ACT and \herschel{} thumbnails in Figure~\ref{fig:thumbnail}; the SMA contours are overlain on the Spitzer, VHS, and PS1 thumbnails (green contours); and the total SMA-derived flux density for this system is plotted as the green star on the SED.
The nineteen SMA-derived flux densities (green points) are included in the SEDs of Figure~\ref{fig:combination}, which
provides a total flux density comparison for the SMA derived values and those from ACT and the best-fitting SED. 
With this comparison and the higher resolution SMA imaging, we infer additional information about the ACT DSFGs and, in particular, investigate whether the source may be composed of multiple components, especially as inferred from multiplicity in the PCAT reduction of the \herschel{} data. 
We detail these findings in Section~\ref{sec:SMAcomparison}.
SMA flux densities are reported in Table~\ref{tab:smacatalog}.

\subsection{SED Model}
\label{sec:modeling}

The primary source of far-infrared radiation in DSFGs is the absorption of starlight by dust, which produces a thermal emission spectrum. 
As such, the optical depth $\tau$ primarily reflects the absorptivity of the medium, and we can model the SED with a MBB spectrum \citep{Su17}. The photon escape probability is:
\begin{equation}
P=1-e^{-\tau}.
\end{equation}
Here, the optical depth $\tau$ is given by
\begin{equation}\tau=\kappa_0 \left (\frac{\nu_r}{\nu_0} \right )^2 \frac{M_d}{\pi \left (\frac{d}{2} \right )^2}\end{equation}
where $\kappa_0=1.5 \mathrm{\ cm^2 \ g^{-1}}$, $c/\nu_0 = 850 \mathrm{\ \mu m}$, $M_d$ is the dust mass, $d$ is the dust emission region diameter and $\nu_r=(1+z)\nu$ is the rest frame frequency \citep{Weingartner01,Dunne03}.

With these, the flux density at temperature $T$ is given by,
\begin{equation}
S(\nu)=\pi \left (\frac{d}{2D_A} \right )^2 (1-e^{-\tau}) \frac{2h}{c} \frac{\nu^3}{e^{h \nu (1+z)/k_B T} - 1},
\end{equation}
which is a model for a fixed dust temperature, where D$_A$ is the angular diameter distance, h is Planck's constant, and k$_B$ is the Boltzmann constant. 
Given a temperature distribution $f(T)$, one may generalize this to
\begin{equation}
S_{\mathrm{multi}}(\nu)=\int_0^{+\infty} f(T) S_\nu(\nu, T) \; dT.
\end{equation}
For the majority of the ACT-selected DSFGs, the peaks of the dust spectra are well sampled by the SPIRE data points; thus, we opt for a power-law description of the dust temperature because it better represents 
the Wien's tail of the modified black body spectrum in comparison to a single-temperature model \citep{Kovacs10}.
We take $f(T)\propto T^{-\gamma}$ with an unknown lower cutoff $T_c$, yielding
\begin{equation}
S_{\mathrm{multi}}(\nu)=(\gamma -1) T_c^{\gamma - 1} \int_{T_c}^{+\infty} S_\nu (\nu,T)T^{-\gamma} \; dT
\end{equation}
We adopt the typical value of $\gamma=7.0$ for DSFGs \citep{Magnelli12}. 

We fit this model to our data using the affine-invariant Markov Chain Monte Carlo (MCMC) \texttt{emcee} sampler \citep{Foreman-Mackey13}, iterated over 6,000 steps with 16 chains and a burn-in of 500 steps. 
This provided us with posterior distributions for $z, d, M_d$ and $T_c$. 
For five of our sources, we re-run fits with the redshifts fixed to their known spectroscopic values, which yields more precise estimates for the remaining parameters.
For each step in the Markov chain, we computed the posterior probability for the parameters as the product of their prior and a Gaussian likelihood. 
We fit the ACT and ensemble \herschel{} flux densities with a single MBB model. 
The negative log-likelihood function for a given ACT source summed over frequencies $\nu$ is given by:
\begin{equation}
    -\log{\mathcal{L}}\propto \frac{1}{2}\sum_{\nu} \frac{(D(\nu) - S(\nu ; z, T_c, M_d, d))^2}{\sigma^2(\nu)},
\end{equation}
where $D(\nu)$ denotes the observed flux densities and $\sigma (\nu)$ their error. 

We also introduced two useful quantities derived from this model. 
The first is the far-infrared luminosity $L_{\mathrm{IR}}$:
\begin{equation}
\label{eq:LIR}
L_{\mathrm{IR}} = 4 \pi D_L^2 \int_{\nu_1}^{\nu_2} S(\nu ; z, T_c, M_d, d) d\nu
\end{equation}
where $D_L$ is the luminosity distance, and the integral is conventionally taken over 8-1000 $\mu$m in the rest frame. 
The second quantity is $\tau_{100}$, the optical depth at $100\mathrm{\ \mu m}$ in the rest frame, approximately where the DSFG spectra peak.

An unknown number of our DSFGs are lensed, which will consequently magnify some of their properties. 
We therefore actually obtain the apparent FIR luminosity $\mu L_{\mathrm{IR}}$, the apparent dust mass $\mu M_d$, and the apparent effective diameter $\sqrt{\mu}d$, where $\mu$ is the unknown magnification associated with a given galaxy, typically found to be in the range 5-10 for the brightest mm-selected galaxies \citep[e.g.,][]{Reuter20}. 

\subsection{Prior Constraints}
\label{sec:priors}
Due to parameter degeneracies in this model, for example between redshift, temperature, and dust mass, 
we require prior information to obtain reliable fit results. 
As a starting point for our prior distributions, we referenced the SED fits to the 81 sources from the South Pole Telescope \citep{Reuter20}, which are spectroscopically complete and are also flux-limited, albeit at a higher flux limit than our sample with $S_{1.4mm} >$20~mJy at 4.5$\sigma$.
Their sample is comprised of sources at the brightest end of the DSFG luminosity function.
There are also significant differences in our model definitions and DSFG flux density distributions that lead us to modify the prior distributions relative to theirs.

\citet{Reuter20} \citepalias[][]{Reuter20} used a single-temperature MBB spectrum with a fixed spectral slope, $\beta$=2, and fit the SED normalization, dust temperature, and redshift as free parameters (they have a second model in which the redshift is fixed to the spectroscopic value).
They then computed apparent total luminosity from the model results and dust mass from the 345~GHz flux density drawn from their best-fitting model SED curve.  
While the \herschel{}-SPIRE flux densities of our sources are of similar distribution as those from \citetalias{Reuter20}, the lower millimeter-wavelength fluxes from ACT may indicate significant differences in the distribution of physical properties of these two samples. 
A two-sample Kolmogorov-Smirnov test definitively indicates the samples are drawn from different distributions with a p-value of 1.95e-20 (See Figure~\ref{fig:flux_histograms}). 

\begin{figure*}[ht!]
\centering
\includegraphics[width = 6.75in]{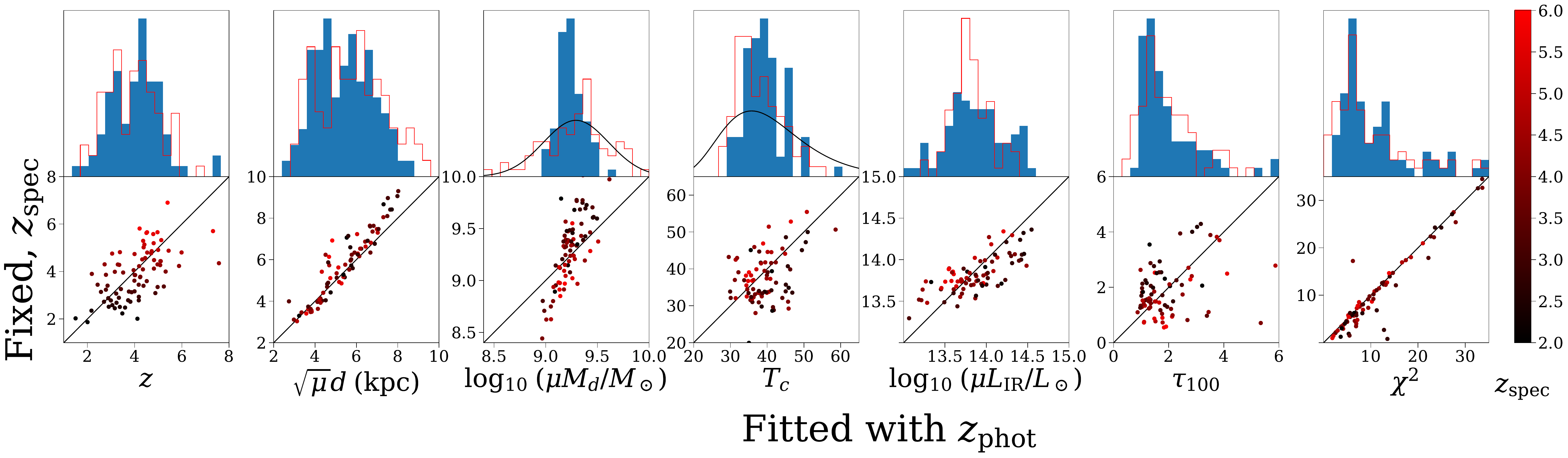}
\caption{Median values of the fit parameters when our model is fit to the flux densities of the \citetalias{Reuter20} sources. 
The red histograms and y-axis of the scatter plots use only the uniform priors on apparent size, apparent dust mass and cutoff temperature, and fixed spectroscopic redshift.
The blue histograms and x-axis of the scatter plots correspond to using the newly defined priors (shown as black curves on the dust mass and temperature histograms) and fitting for redshift. The points are colored by their spectroscopic redshift. }
\label{fig:ReuterAndACT}
\end{figure*}

To determine our priors, we fit the flux densities of the \citetalias{Reuter20} DSFGs from their Table~D1 (from 3mm to 250\micron) with our model as defined in Section~\ref{sec:modeling}, fixing the redshift to the spectroscopic values of their sample to break its degeneracy with the rest of the model parameters. 
We imposed flat priors on size, dust mass, and cutoff temperature: $\sqrt{\mu} d \; (\mathrm{kpc}) \in (0.5, 15.0)$ and $\log_{10} (\mu M_d/M_\odot) \in (7.0, 11.0)$ and $T_c \; (\mathrm{K}) \in (15.0, 60.0)$.
We then defined our prior distributions for dust mass and cutoff temperature using the resulting posterior distributions from these fits. 
For cutoff temperature, we imposed a log-normal prior distribution centered at $T_c = 40.7 \pm 12.2 \; \mathrm{K}$. 
For dust mass, we imposed a normal prior distribution centered at $\log_{10} (\mu M_d/M_\odot) = 9.29 \pm 0.32$ and a strict cutoff of $\log_{10} (\mu M_d/M_\odot) \in (7.0, 12.0)$. 
For redshift and size, we required $z \in (0.5, 12.0)$, $\sqrt{\mu} d \; (\mathrm{kpc}) \in (0.2, 11.0)$.

We tested the effect of imposing these priors by modeling the \citetalias{Reuter20} DSFGs again using these newly defined prior distributions and fitting for photometric redshift. 
In the comparison between these fits and the fits to the \citetalias{Reuter20} flux densities using their spectroscopic redshifts and our model definition without a prior (Figure~\ref{fig:ReuterAndACT}), we find agreement in the fit parameters with correlation coefficients of 0.58, 0.45, 0.75, and 0.94 for redshift, cutoff temperature, dust mass, and size respectively.
Imposing the prior significantly tightens the distributions of cutoff temperature and dust masses. 
This may be due to the multiple degeneracies between dust mass, cutoff temperature, and redshift such that the combination of priors on dust mass and cutoff temperature leads to a tightening of the posterior distribution of dust masses around the peak of the prior.
We further show in Section~\ref{sec:result_params} that knowledge of the spectroscopic redshift significantly tightens the constraints on the other physical parameter estimations.

\section{Results and Discussion}
\label{sec:results}

Here we report the results of fitting the ACT-selected DSFGs with our SED model and compare the inferred average physical properties of this population with those of other millimeter-selected samples. 
We further examine in this section the likelihood that our sources are singular, lensed or inherently bright DSFGs through comparisons of the SMA flux densities and assessment of the PCAT flux component multiplicity.

\subsection{Physical characteristics of ACT DSFGs}
\label{sec:result_params}
We fit our model to the six ACT and SPIRE ensemble flux densities for each of our 71 sources. 
For all sources, we generated best-fitting physical characteristics using the median of the posterior distributions from the MCMC modeling, and we report these along with their 16th and 84th percentile values and the $\chi^2$ statistic in Table~\ref{tab:fitparams}.
We find reasonable goodness of fit for the majority of the SEDs, but there are some outliers. 
We report the $\chi^2$ statistic for the model determined by the median of each parameter. 
In some cases, the median parameter model curve is a less good fit to the data, as compared to the parameter values at the location of the maximum likelihood, which often occurs when the posterior probability distributions deviate from normal. In particular, the apparent size is poorly constrained for a minority of our sources, but we conclude that these deviations do not change the statistical inferences we make herein.

From the distribution of parameter medians for each individual source fit, we find the population median, 16th and 84th percentiles of each parameter to be $z_\mathrm{phot}=3.3^{+0.7}_{-0.6}$, $\sqrt{\mu} d=5.2^{+0.9}_{-2.4} \; \mathrm{kpc}$, $\log_{10} (\mu M_d/M_\odot)=9.14^{+0.12}_{-0.04}$ and $T_c=35.6^{+4.8}_{-1.6} \; \mathrm{K}$. 
The apparent FIR luminosity and optical depth at 100 $\micron$ are $\log_{10} (\mu L_\mathrm{IR}/L_\odot)=13.6^{+0.2}_{-0.3}$ and $\tau_{100}=1.5^{+4.5}_{-0.4}$.

\begin{figure*}[ht!]
\epsscale{1.0}
\plotone{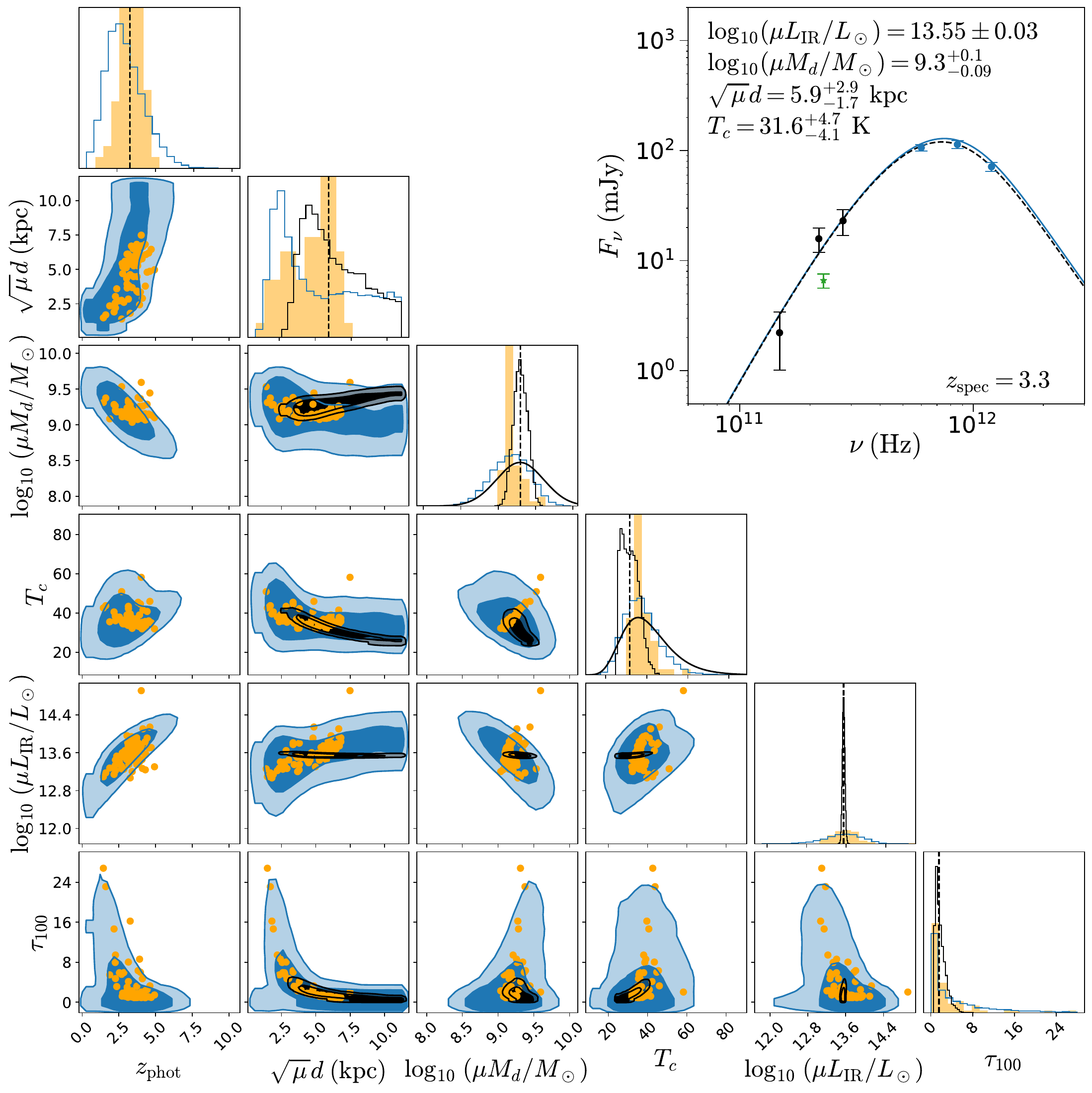}
\caption{Posterior probability distributions of SED model parameters for ACT-S J010729+000114. Blue and black contours (histograms) are 2D (1D) marginalized distributions derived by fitting photometric redshift and by fixing the spectroscopic redshift, respectively. Dark and light contours indicate 68\% and 95\% confidence regions. The vertical black dashed line indicates the median values. The prior distributions for cutoff temperature and dust mass are shown as black curves on the histograms. The plot in the top right shows the SED for this source with the same plotting conventions as Figure \ref{fig:combination}. The median parameter values are indicated on the upper left of the SED and the spectroscopic redshift on the bottom right. The orange points and filled histograms show the median results for all 71 sources. 
}
\label{fig:MCMCcorner}
\end{figure*}

Figure~\ref{fig:MCMCcorner} shows the posterior distributions for the galaxy ACT-S J010729+00011 as blue contours and the median values for all other ensemble runs in orange.
The black contours show the results of the model using the spectroscopic redshift, $z_{spec}=3.33$, for this source. 
The photometric redshift inferred from the fit is $z_{phot}=4.0^{+1.3}_{-1.0}$.
Removing the degeneracy with redshift allows for much tighter constraints on cutoff temperature and dust mass and, therefore, apparent luminosity. 
We see a positive correlation between apparent luminosity and photometric redshift, and a negative correlation with apparent dust mass. 
Fixing the redshift to its spectroscopic value causes the luminosity distribution to be symmetric with respect to changes in the fit parameters. 
This symmetry and the tighter constraint of luminosity with fixed redshift may indicate that redshift has a much greater influence on the apparent luminosity than the other parameters.

\subsection{Comparison with other samples and possible biases}
\begin{figure*}[h!]
\epsscale{1}
\plotone{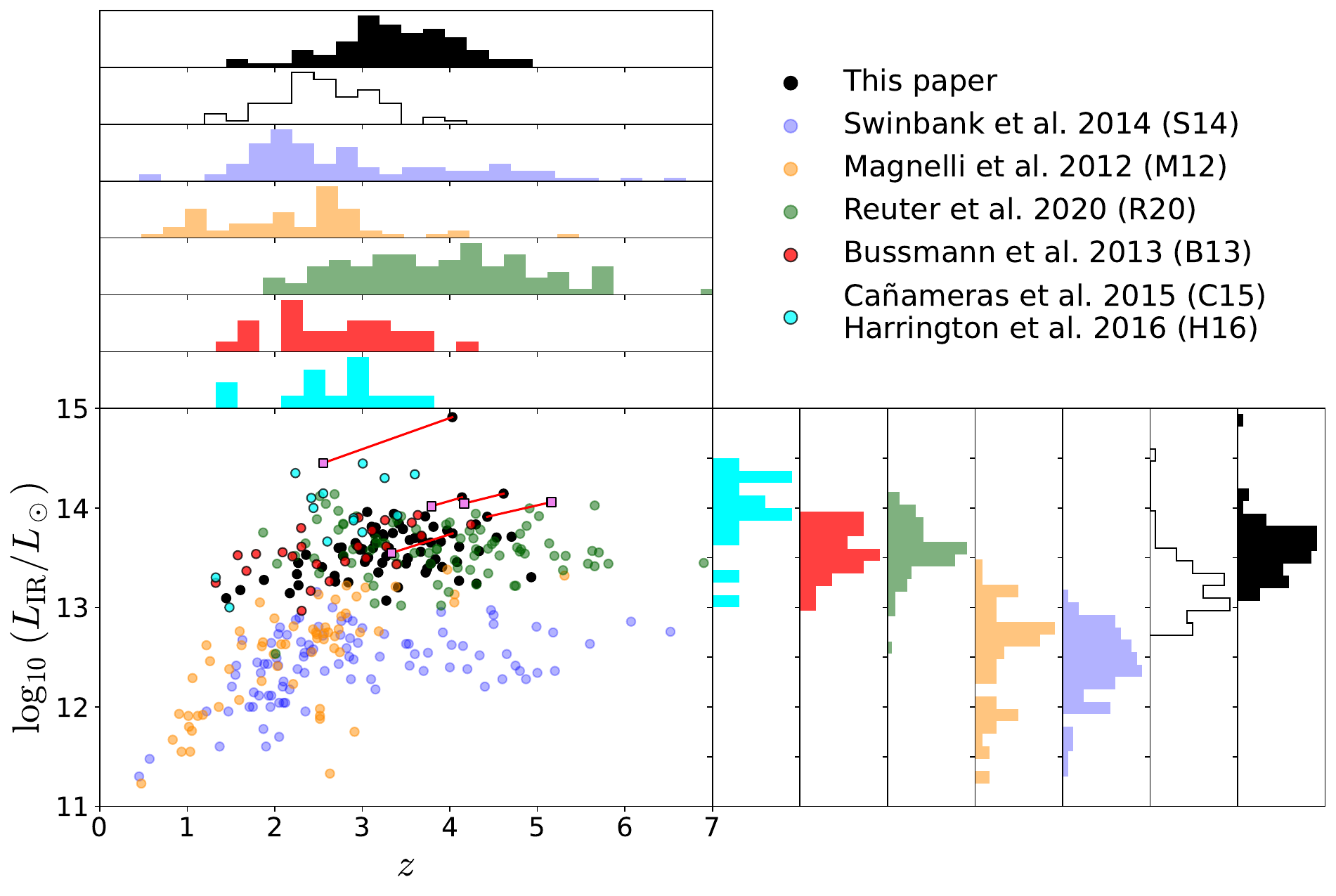}
\caption{A comparison of redshifts and apparent or intrinsic FIR luminosities from various millimeter and submillimeter-selected samples of DSFGs. 
The sources from \citetalias{Reuter20}, B13 and C15/H16 are lensed, while those from \citetalias{Magnelli12}  and \citetalias{Swinbank14} are de-magnified or unlensed. 
\citetalias{Reuter20} sources all have spectroscopic redshifts.
Histograms for the apparent luminosities and redshifts from each sample are provided on the lower right and upper left, respectively. 
ACT and \citetalias{Magnelli12}  use a power-law temperature distribution, while the others use a single-temperature model. 
The hollow black histograms correspond to ACT fits using a single-temperature model. 
ACT results using a fixed spectroscopic redshift are shown with violet squares, and are connected to their photometric redshift results by red lines.}
\label{fig:LvsZ}
\end{figure*}

We calculated the apparent far-IR luminosity distribution for each source using equation~\ref{eq:LIR}, and compare the distribution of medians as a function of redshift with that from other samples.
$L_{IR}$ is a parameter of interest as it corresponds to the obscured star formation rate of the galaxy, which in turn is statistically significant for this population as it is responsible for the bulk of the cosmic star formation rate density as least between $z=1-4$ \citep{hall18, zava21}.

In Figure~\ref{fig:LvsZ}, we compare the redshifts determined from the ensemble fits with their apparent (or intrinsic) FIR luminosity. 
We also show DSFGs from other samples, namely the ALMA Laboca ECDFS Submillimeter Survey \citep[ALESS;][S14]{Swinbank14}, \cite{Magnelli12} (M12, \herschel{} PACS/SPIRE $100–500$ $\mu$m), the South Pole Telescope \citep[SPT;][R20]{Reuter20}, \cite{Bussmann13} (\herschel{} SPIRE $250 \ \mu$m, $350 \ \mu$m, and $500$ $\mu$m), and the overlapping \cite{Canameras15} and \cite{Harrington16} samples (Planck-\herschel{} SPIRE 250-850 $\mu$m). 
We use only the parameters provided in those papers for comparison.

The galaxies in the \citetalias{Swinbank14} and \citetalias{Magnelli12} samples are unlensed or de-magnified, and therefore we use their intrinsic luminosities, which are noticeably lower than the apparent luminosities from the other papers, as expected as the others are apparent luminosities likely amplified by lensing or multiple blended sources. 
The \citetalias{Swinbank14} sample has photometric redshift estimates for 99 submillimeter galaxies (SMGs), while \citetalias{Magnelli12}  is made up of 61 sources with spectroscopically determined redshifts. 

As was indicated in Figure~\ref{fig:MCMCcorner}, there is a degeneracy (in general, a positive correlation) between redshift and apparent luminosity, emphasizing that precise redshift measurements have an important role in our ability to extract physical information about the DSFGs. 
We see this again in Figure~\ref{fig:LvsZ} where we include additional data points for our five sources that have spectroscopic redshifts measurements (pink squares).
In all but one source, the spectroscopic redshift differs from the photo-z by $\leq$0.6$\sigma$, with J020941+001557 (``the Red Radio Ring") being the exception with a spectroscopic redshift that is lower than the fit result by 1.5$\sigma$. 
Using the spectroscopic redshift values reduces the apparent luminosity by $\Delta \log{(L_{IR}/L_\odot)}/\Delta z=0.27 \pm 0.05$ for four out of five of the sources (the redshifts of these sources are smaller than the model-derived photometric redshifts), and increases the LIR (with increasing spectroscopic redshift) for the fifth source by a similar value, $\Delta \log{(L_{IR}/L_\odot)}/\Delta z=0.22$.

The spectroscopic redshifts of the \citetalias{Reuter20} sources (Figure~\ref{fig:LvsZ}, green points/histograms) have a wider distribution and higher mean redshift compared to the other samples, including ours. 
This discrepancy can be explained for the \herschel{} and ALESS samples by the SPT’s mm-wave selection and large survey area, respectively; however, this does not explain SPT’s disagreement with ACT’s results. 
When we use a single-temperature blackbody model (Figure~\ref{fig:LvsZ}, empty black histograms) instead of our power law temperature distribution model (Figure~\ref{fig:LvsZ}, filled black histograms), our fitting results suggest an even larger shift to lower redshifts, and the median of the apparent luminosity distribution shifts lower by 1.4$\sigma$ in comparison to our fiducial model. 
Similarly, the median apparent luminosity of our fiducial model fits is 1$\sigma$ lower in comparison to that of the \citetalias{Reuter20} sources with the same model, which could be indicative of our sources having overall weaker lensing or a higher fraction of unlensed DSFGs in our sample.
Moreover, our use of the priors for apparent dust mass and cutoff temperature from \citetalias{Reuter20} might be expected to bias our derived redshifts to higher values, but we instead see the lower median in the distribution. 
This could be further evidence that this ACT-selected DSFG sample is tracing a different population of DSFGs.

The apparent luminosity comparison must be considered in conjunction with other degeneracies of the model parameters. 
Our fiducial model produces systematically lower dust masses than those reported in \citetalias{Reuter20}, which are computed from the 345~GHz flux density of their best-fitting model curve. 
Using our model to fit the \citetalias{Reuter20} flux densities and spectroscopic redshifts, we find the median of our dust mass distribution is 2$\sigma$ below the median of the \citetalias{Reuter20} dust mass distribution. 
As discussed in Section~\ref{sec:priors}, we use these fits to define the prior constraint on dust mass, and cutoff temperature, which needs to be considered in our broad interpretation of the parameter constraints from our model.
Our lower apparent dust mass distribution in comparison to the \citetalias{Reuter20} results using the same model may indicate a difference in the physical properties of our DSFGs selected with a lower flux limit. 
This result could be further evidence of overall weaker lensing or the fact that our sample is constructed of a higher fraction of unlensed sources.


\subsection{Comparison with SMA fluxes}
\label{sec:SMAcomparison}

For each of the nineteen sources, we analyzed the difference between the SMA flux densities and the value of the ensemble model at the same frequency. 
Six of the nineteen sources' SMA flux densities -- J001133-001835, J003337+000353, J003929+002422, J004532–000127, J005847-010025,  J011640–000457 -- agree within uncertainty  with the flux densities extracted from the best-fit spectrum to the ACT and ensemble \herschel{} fluxes, with two of them measuring slightly higher by 2.2$\sigma$ and 1.5$\sigma$. 
Five of these sources (excluding J004532-000127) provide no indication of missing flux or additional sources in the SMA data, but in each case the source is partially resolved by the SMA synthesized beam. 
We consider their potential to be blended even in the SMA data in addition to possible lensing features. 
In particular,
\begin{itemize}
    \item J001133-001835 is partially resolved from the clean beam, which has size 4.32"$\times$2.52". It has a morphology that is not well fit by a 2D Gaussian, but there is no evidence for a lens in the optical or near-infrared imaging. 
    
    \item J003337+000353 is partially resolved beyond the size of the clean beam size, which is 9.65"$\times$3.09". The total flux density extracted from the SMA data is 2.2$\sigma$ above that predicted by the SED fit, and thus still in agreement.
    
    \item J003929+002422 is partially resolved beyond the size of the clean beam, which is 4.1"$\times$2.52".  The total flux density extracted from the SMA data is 1.5$\sigma$ above that predicted by the SED fit. There is a potential lens seen in the near infrared

    \item J005847-010025 is partially resolved, and it was observed in the sub-compact configuration with the SMA and so has a clean beam size of 7.7"$\times$4.14".

    \item J011640-000457 is also partially resolved from the 3.89"$\times$3.26" clean beam, and it exhibits non-gaussian morphology, which likely indicates lensing features. This target was also observed with the Northern Extended Millimetre Array to detect the CO(4--3) emission line, and it was determined to lie at $z=3.7908$ and has a strong lensing morphology in these data with a synthesized beam of 1.37"$\times$0.74" \citep{Rivera19}.  The lens candidate is visible in the Spitzer, VHS, and Pan-STARRS1 images and was identified in \citetalias{Gralla20} as an SDSS galaxy at $z=0.45$.
    
    \item The SMA data for J004532–000127 capture all of the flux as expected by the ensemble ACT and \herschel{} fit when including the multiple sources that lie within $\sim$10" (Figure~\ref{fig:thumbnail}, green contours). While there is a source in the near-IR and optical images that suggest this could be a lens, the mass required to generate a $>$10" lensed arc is that of a cluster, not via galaxy-galaxy lensing. The projected distance to the galaxy is also too large, and the clumpy nature of the SMA detections also suggest this is not a lensed arc. 

\end{itemize}

\begin{figure}
    \centering
    \includegraphics[width=\linewidth]{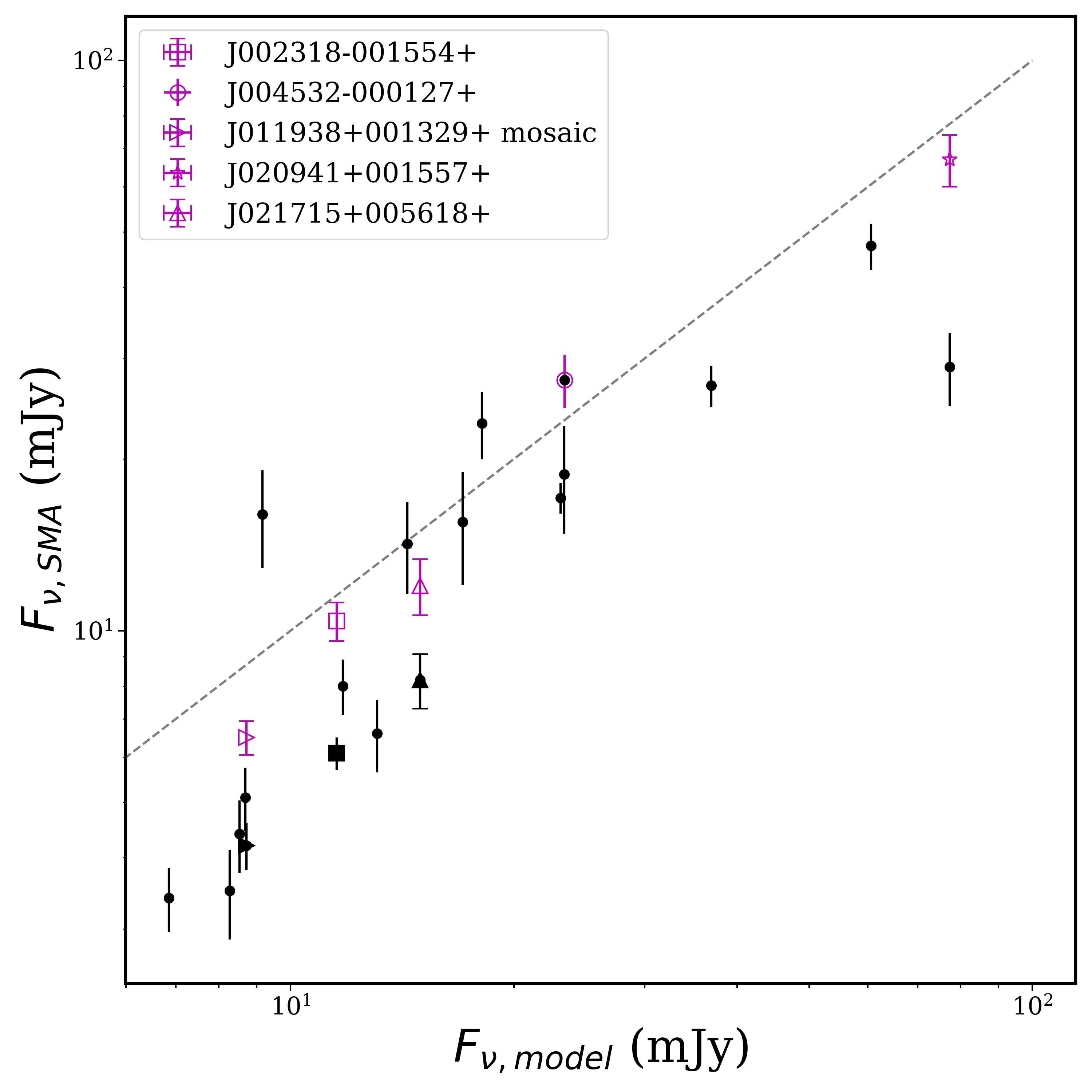}
    \caption{SMA flux density versus the ensemble model flux density determined from the median SED. For all sources, the filled black symbols are the flux densities of the source closest to the ACT source. Sources with multiple SMA detections are listed in the legend. For these, the matched unfilled magenta symbols include the recovered flux from the additional source(s) in the SMA data.} 
    \label{fig:smacompare}
\end{figure}

Of our nineteen sources imaged at higher resolution, we find that thirteen SMA flux densities lie below what is expected if all of the ACT and \herschel{} flux is coming from just one source. 
In addition to J004532–000127, we detect an additional source in four other SMA observations.
Figure~\ref{fig:smacompare} plots the SMA flux density versus the ensemble model flux density determined from the median SED.
For sources with multiple SMA detections, the unfilled magenta symbols are the flux densities recovered from all sources, and the matching filled black symbols are the flux density of the source closest to the ACT source (this is not always the brightest source). 
One of these secondary sources was detected only after additional follow-up with the SMA observing with a seven point mosaic pattern around the ACT source center.
These SMA sources are discussed further in the following section.

One concern with interferometric observations is that flux on large scales can be resolved out of the images. All of our observations taken with the SMA in either compact or sub-compact configuration such that the maximum recoverable scale is $\sim$28", but that is not without loss of flux. We compute the flux loss as a function of angular extent of an approximately 2D Gaussian source \citep{wiln94}.
In the configurations for which our sources were observed, the SMA can recover at least 90\% of the flux of sources with angular extent $\leq$5.4".
We do not expect to have extended emission beyond this scale (which is 42~kpc at z=3) from a single source. An exceptional case would be a strong lens by a massive group or cluster, which would be non-Gaussian in shape. We estimate the flux loss for an extreme case of a 10" lensed arc by simulating the observations using our SMA antenna configurations. We find that we still recover 82.5\% of the flux for an 8~mJy source (the lower limit of the ACT detections). This is a highly unlikely scenario as it requires a very massive lens, which we do not see in the optical or near-IR images of our fields. 

\subsection{Multiplicity and Protoclusters}
\label{sec:multi}

We expect a fraction of our sources to be unlensed systems based on the millimeter flux limit, and other findings for DSFGs in overdense environments \citep{Ma19, wang21, cox23, quir24}. 
\citetalias{Reuter20} reported that two of their DSFGs are actually multiple systems, as seen in the ALMA images. 
One of these resolves into two sources at $z=6.9$ and the other is one of the most overdense protoclusters known at $z=4.3$ \citep{oteo18}.
\citet{wang21} published a follow-up study of APEX/LABOCA observations of nine of SPT protocluster candidate fields, covering an area of 1300~deg$^2$ and identified 98 sources in the combined area.
\citet{hodg13} published the source catalog from the ALESS survey, finding 32-50\% of the targeted SMGs were resolved into multiple sources. 
\citet{quir24} report that 20\% of their red-\herschel{} targets that are detected by ALMA resolve into multiple systems, and find further evidence that below a flux limit of 13~mJy at 1.3~mm, DSFGs are likely to be unlensed (a similar result to the far-IR constraint from \citealt{Negrello10}). 

Here we discuss the validity of the PCAT flux components as potential indicators of multiple galaxies by considering the flux densities and multiplicity of the nineteen sources with SMA observations.
The PCAT multiplicity is greater than one for 58 of the 71 DSFGs ($\sim$82\%).
Due to the \herschel{} resolution and the instrument noise limit, this fraction is expected to be higher than the quoted fractions defined with the ALMA de-blending radii, and also with respect to the fraction of true physically associated multiples.
Simulations to characterize the contamination from field galaxies, or the performance of PCAT on mock cases with known multiplicities, are beyond the scope of this work, but are what would be required to better connect the "number of flux components" to intrinsic multiplicity of these systems.

We use the results of the SMA imaging to comment on multiplicity.
Of the six sources for which the total SMA flux density agrees with that of the median SED, the PCAT multiplicity is greater than one for five of them (all excluding J001133-001835). 
The PCAT multiplicity for J004532-000127 (Figure~\ref{fig:thumbnail})is two, while the SMA data reveal at least three sources within approximately 15". 
Given the density of this environment, it is not expected that PCAT could differentiate all of these sources.
As mentioned in the previous section, with the exception of J011640-000457, we cannot say for certain whether the other four sources are multiples due to the size of the SMA beam, though J003929+000353 is also indicated to be lensed.

We know of at least one other case in which the PCAT multiplicity is larger than the number of flux components that are likely in the field, and that is for J020941+001557, (the ``Red Radio Ring", \citealt{Geach15}). 
This is a known lensed DSFG at $z=2.55$. 
This galaxy has been observed at radio and millimeter wavelengths at $<$1" resolution by VLA and $\sim$11" and 15" resolution by SCUBA-2 \citep{Geach15} and $\sim$1" resolution by NOEMA \citep{Rive19}, and there is a known continuum flux decrement as compared to the ACT flux. 
With the larger field of view of the SMA, we recover an additional millimeter-bright source in the field at $\sim$24" distant from J020941+001557. 
With the additional contribution from this second source in the SMA field, we almost recover the flux density expected from the median SED, though the total is still less than expected by 1.5$\sigma$. 
The PCAT multiplicity for this source, however, is six.
The $\chi^2_{red}$ for the PCAT run on this source is higher than average, and this poor fit and over-prediction of source components is likely due to either errors on the point spread function, which are more pronounced for the brightest sources, or to inadequacies of the point source description for the blended emission. 
J020941+001557 is brighter than any other source in our sample by a factor of four, and by much more than that in most cases. 
From the distribution of $\chi^2_{red}$ and source flux densities we do not expect this type of over-prediction to be common.
Any conclusions drawn about this source's flux and multiplicity come from the SMA observations.

We find fourteen out of the nineteen SMA-observed sources show evidence for multiplicity as indicated by either detecting additional sources or an SMA flux density that is significantly ($\geq 3\sigma$) below the ensemble SED. 
The PCAT multiplicity for eleven out of these fourteen sources is greater than one; the remaining three have multiplicity of one within the 30" radius inside of which we extract that information.
Other than J004532-000127 and J020941+001557, three more SMA observations -- J002318-001554, J011938+001329, and J021715+005618 -- reveal at least one additional source in the field. 
In the case of J011938+001329, PCAT assigns a multiplicity of one; the second SMA-detected source was only confirmed after following up the initial observations with a seven point mosaic pattern surrounding the ACT location. 
The additional source is 14" to the NE of the more central source, which itself is offset from the ACT location by 16". 
This not only puts the second source near the edge of the expected radius for contribution to the ACT flux density, but its flux as measured by the SMA is 2.3~mJy, well below the ACT 5$\sigma$ flux limit. 
Nonetheless, this source would still contribute to the total blended flux density of the ACT source, and provide supporting evidence for the need for deeper high resolution observations.
A similar scenario was seen in \citet{cair23} in which there were initially no SMA detections of several bright \herschel{} SPIRE sources, but were then resolved into fainter multiple components upon deeper observations.

In each of these cases, we do not have redshift information for the sources, so cannot say whether these are associated galaxies or if they are at differing redshifts. 
For the case of J004532-000127, we infer that these galaxies are associated based on their proximity and even possible indicators of interaction in the SMA data, but this is speculation until we can secure redshift information. 
We can also possibly infer from the shape of the SEDs of the brightest flux component from PCAT whether we might expect there to be interloping flux from higher or lower redshift sources by examining where the brightest flux component SED peaks relative to the ensemble flux, but we recommend caution and additional high-resolution follow-up for interpreting these interesting cases.

\section{Summary and Conclusions}
\label{sec:conclusion}

We model the millimeter through far-infrared spectral energy distributions (SEDs) of 71 ACT-selected DSFGs using ACT-derived de-boosted fluxes from \citetalias{Gralla20} and the ensemble fluxes derived using the PCAT algorithm from the \herschel{} SPIRE cutouts around the ACT source locations. 
PCAT uses forward modeling to perform a probabilistic reconstruction of the \herschel{} cutouts into flux components, reliably resulting in between 1-4 components within a 30" radius around the ACT source centers, with the exception of one outlier source. 
Owing to the lower flux-limit of the 1.4~mm data of 8~mJy, it is expected that a portion of the ACT-selected DSFGs are unlensed and our results herein support that expectation. 
Our primary findings are as follows:
\begin{itemize}
\item Our sample of DSFGs have similar apparent luminosities as other mm-selected sources, but we exercise caution in interpreting these results owing to the possibility that the ensemble flux densities are actually multiple source components contributing to the ACT and \herschel{} points.

\item The median photometric redshift of our sources is $z_{phot} = 3.3^{+0.7}_{-0.6}$. Independent of the dust-temperature model and despite priors based on the higher-$z$ spectroscopic sample from SPT, this redshift distribution is below that of SPT, implying that the primarily-lensed bright end of the 1.4~mm-selected DSFG population \citepalias{Reuter20} probes an overlapping but different star-formation epoch from surveys at similar wavelengths but with lower flux-density limits that extend into the unlensed population (this work). 
 
\item Our dust masses are lower than inferred by \citetalias{Reuter20}. This result may also be interpreted as an indicator that this ACT-selected sample contains a higher fraction of unlensed sources or generally less-massive DSFGs. We exercise caution in that degeneracies between dust mass, temperature, and redshift in the model definition make it difficult to fully interpret this result. 

\item The PCAT diagnostics indicate that 82\% of our sample could be composed of multiple flux components. This decomposition provides a basis for further high resolution follow-up of the sample. 
Images of the nineteen sources observed with the SMA indicate that at least fourteen (74\%) are likely unlensed and possible multiples, but the resolution of many of these data is not high enough nor are they necessarily sensitive enough observations to unambiguously determine whether there are additional sources in the field. 
We unambiguously determine that five of our sources are comprised of at least two sources. 
PCAT indicates a high probability of multiplicity values greater than one for eleven of the fourteen sources for which the SMA data show multiple source detections or indicate missing flux relative to the ensemble model. 

\end{itemize}

We conclude that this sample of DSFGs is likely tracing an intermediate luminosity/mass range of the overall DSFG population. 
Evidence herein suggests that this sample contains a higher fraction of unlensed galaxies and possibly a higher fraction of multiplicity in comparison to the high-flux/high-luminosity DSFGs identified in \citetalias{Reuter20}, the closest millimeter-selected survey sample to which we can compare.
The \herschel{} maps in these regions (HerS and HeLMS) are instrument noise dominated, which limits the effectiveness for the PCAT algorithm to unambiguously de-blend the data into multiple flux components. 

Continued identification and characterization of the point sources in the ACT maps serves as a valuable tool for identifying sources/regions for higher-resolution follow-up.
While existing single dish facilities, such as the Large Millimeter Telescope and James Clerk Maxwell Telescope, continue to push the envelope on detailed follow-up of DSFG samples, our SMA sample in this study sets the sensitivity limits deeper than previously thought necessary.
The next generation of far-infrared and millimeter wavelength telescopes and instruments, such as the proposed Atacama Large Aperture Submillimeter Telescope (AtLAST, \citealt{klaa20}) or the PRobe far-Infrared Mission for Astrophysics (PRIMA), will allow us to more efficiently map wide-fields with simultaneously spatial and spectral separation \citep{beth24,vank24}. 
Additional high resolution follow-up, ideally with sensitive enough observations to observe CO emission and determine source redshifts is required in order to fully characterize the physical properties of this sample of millimeter flux-limited DSFGs.



\begin{acknowledgments}
    K. R. Hall acknowledges the Smithsonian Institution for funding her as a Submillimeter Array Fellow while carrying out this work. The Submillimeter Array is a joint project between the Smithsonian Astrophysical Observatory and the Academia Sinica Institute of Astronomy and Astrophysics and is funded by the Smithsonian Institution and the Academia Sinica. The authors wish to express immense gratitude for the use of the Mauna Kea observing site, which is one of the most revered places in Hawai'i. We thank Megan Gralla for her preparation of the catalog and guidance on the modeling, and we thank Ryan Wills for creating the \herschel{} maps used for this analysis. We are very grateful to the anonymous reviewer for their time and careful assessment of this manuscript, which led to improvements in clarity and rigor. 
\end{acknowledgments}

\bibliography{main}
\bibliographystyle{aasjournal}

\appendix
\section{Source properties}
\counterwithin{figure}{section}
Here we provide a comparison of the millimeter and far-infrared flux densities of our sources with those from SPT sources from \citetalias{Reuter20} and ALESS sources from \citetalias{Swinbank14}. 
The \citetalias{Reuter20} sources are almost all lensed DSFGs on the brightest end of the luminosity function, while the \citetalias{Swinbank14} sources are unlensed.
Our sample of ACT-selected DSFGs reach flux density at 2~mm and 1.4~mm that are lower than similar low-resolution SPT-selected sources.
We expected a non-negligible fraction of sources at these flux limits to be unlensed and possibly multiples. 

\begin{figure*}[h!]
\includegraphics[width=6.5in]{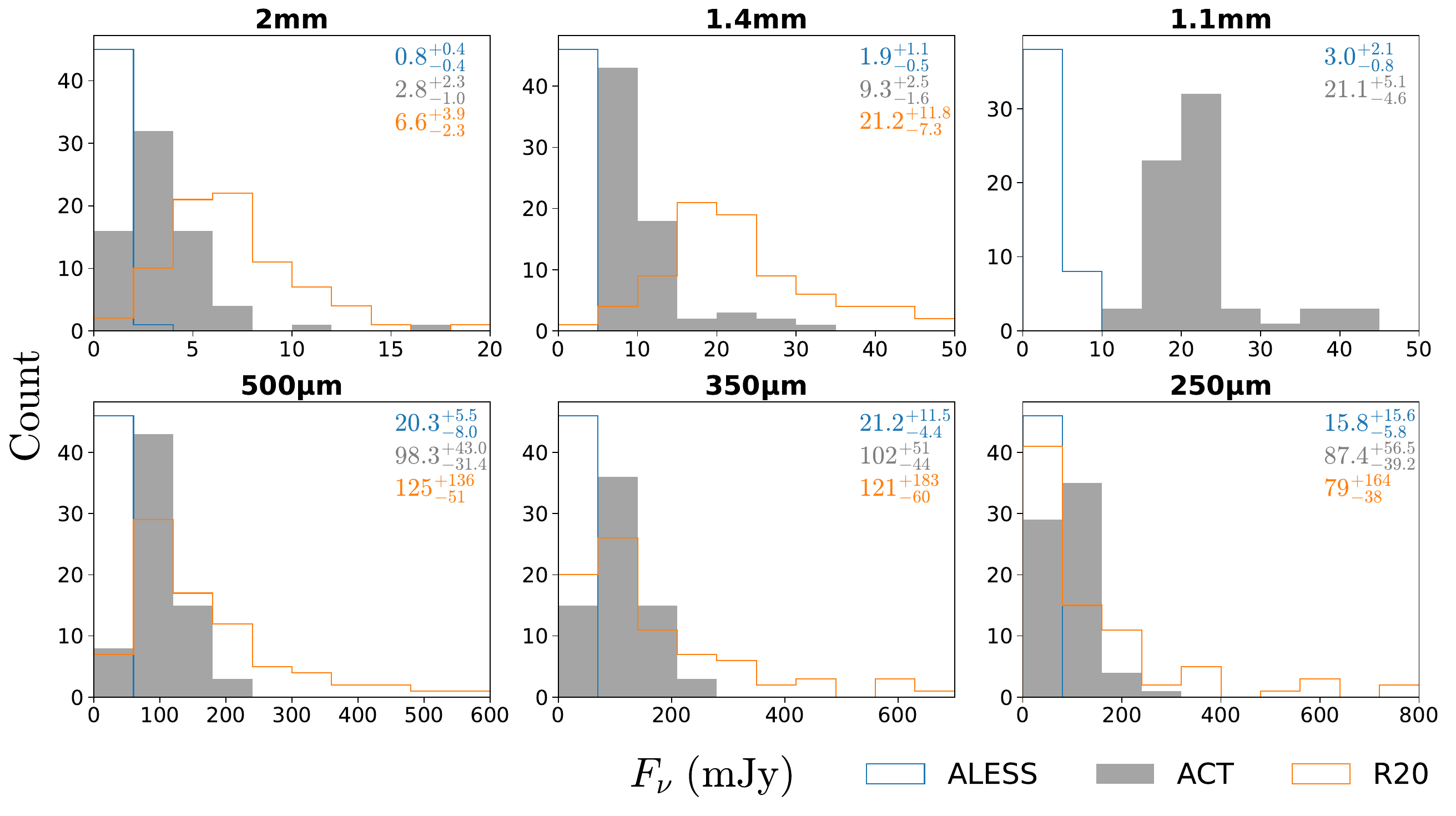}
\caption{Comparison of flux densities (in mJy) across different wavelengths for our ensembles (gray), \citetalias{Reuter20} sources (orange) and ALESS sources (blue). 
For the ALESS sources, we use the ALMA 870~$\mu m$ and \herschel{} SPIRE flux densities from \citet{Swinbank14}; using these submillimeter flux densities, we modeled the SEDs of the ALMA sources, and report the 2~mm, 1.4~mm and 1.1~mm flux densities drawn from the best fit.}
\label{fig:flux_histograms}
\end{figure*}


\begin{longtable}{c c c c c c c c c}
 \caption{ACT source IDs, \herschel{} footprint, PCAT multiplicity, and ensemble fluxes} \\
 \hline
 ACT-S ID & Footprint & Multiplicity & $S_{\mathrm{2000}}$ & $S_{\mathrm{1400}}$ & $S_{\mathrm{1080}}$ & $S_{\mathrm{500}}$ & $S_{\mathrm{350}}$ & $S_{\mathrm{250}}$ \\ (J2000) & & (Median) & (mJy) & (mJy) & (mJy) & (mJy) & (mJy) & (mJy)\\ \hline\endfirsthead
 \caption{(Continued)}\\
 \hline
 ACT-S ID & Footprint & Multiplicity & $S_{\mathrm{2000}}$ & $S_{\mathrm{1400}}$ & $S_{\mathrm{1080}}$ & $S_{\mathrm{500}}$ & $S_{\mathrm{350}}$ & $S_{\mathrm{250}}$ \\ (J2000) & & (Median) & (mJy) & (mJy) & (mJy) & (mJy) & (mJy) & (mJy)\\ 
 \hline
 \endhead
 \label{tab:catalog}
 001133$-$001835 & HELMS & 1 & $5.3 \pm 2.0$ & $20.0 \pm 3.5$ & $24.4 \pm 6.7$ & $99 \pm 10$ & $127.4 \pm 7.0$ & $98.7 \pm 8.9$\\
 001855$-$010649 & HELMS & 1 & $1.9 \pm 1.2$ & $9.5 \pm 3.7$ & $19.1 \pm 5.4$ & $78.2 \pm 8.6$ & $84.5 \pm 6.3$ & $48.3 \pm 5.2$\\
 002147+002445 & HELMS & 2 & $3.8 \pm 1.8$ & $8.4 \pm 3.3$ & $23.9 \pm 5.6$ & $72 \pm 21$ & $72 \pm 17$ & $36.0 \pm 9.2$\\
 002220$-$015523 & HELMS & 2 & $5.0 \pm 1.9$ & $26.3 \pm 3.8$ & $41.9 \pm 5.2$ & $134.0 \pm 6.7$ & $124.9 \pm 6.3$ & $80.4 \pm 6.5$\\
 002318$-$001554 & HELMS & 2 & $3.5 \pm 1.7$ & $11.1 \pm 4.3$ & $20.2 \pm 5.3$ & $93.4 \pm 7.2$ & $94.7 \pm 7.3$ & $67.9 \pm 7.2$\\
 002741$-$011651 & HELMS & 1 & $1.7 \pm 1.0$ & $10.7 \pm 4.1$ & $17.0 \pm 4.9$ & $68.3 \pm 7.6$ & $53.1 \pm 4.9$ & $35.1 \pm 5.2$\\
 003337+000353 & HELMS & 2 & $2.7 \pm 1.5$ & $8.8 \pm 3.5$ & $22.7 \pm 5.3$ & $66.9 \pm 4.9$ & $78.9 \pm 7.6$ & $51.3 \pm 8.1$\\
 003648$-$002052 & HELMS & 3 & $2.0 \pm 1.2$ & $8.4 \pm 3.3$ & $20.3 \pm 5.3$ & $112.9 \pm 8.2$ & $134.4 \pm 7.7$ & $150 \pm 10$\\
 003757$-$010621 & HELMS & 1 & $5.7 \pm 1.8$ & $11.1 \pm 4.3$ & $21.3 \pm 5.3$ & $121.1 \pm 7.6$ & $153.4 \pm 5.5$ & $124.4 \pm 7.1$\\
 003814$-$002255 & HELMS & 2 & $7.1 \pm 1.8$ & $23.5 \pm 3.3$ & $36.9 \pm 5.3$ & $118 \pm 13$ & $122 \pm 11$ & $96 \pm 13$\\
 003929+002422 & HELMS & 2 & $5.1 \pm 1.7$ & $17.7 \pm 4.2$ & $29.9 \pm 5.3$ & $173.7 \pm 8.1$ & $155.1 \pm 6.9$ & $159.5 \pm 7.3$\\
 003943$-$003952 & HELMS & 2 & $2.0 \pm 1.2$ & $8.6 \pm 3.4$ & $26.2 \pm 5.4$ & $91 \pm 16$ & $82 \pm 11$ & $71 \pm 20$\\
 004033+000228 & HELMS & 2 & $3.1 \pm 1.6$ & $8.4 \pm 3.3$ & $15.7 \pm 4.9$ & $67 \pm 16$ & $48 \pm 17$ & $49 \pm 17$\\
 004338$-$005320 & HELMS & 2 & $1.9 \pm 1.2$ & $10.3 \pm 4.0$ & $21.1 \pm 5.3$ & $70.9 \pm 6.4$ & $87.9 \pm 7.9$ & $68.0 \pm 9.7$\\
 004410+011818 & HELMS & 2 & $11.7 \pm 1.8$ & $34.7 \pm 4.2$ & $70.9 \pm 5.0$ & $199 \pm 11$ & $189.5 \pm 8.9$ & $124.7 \pm 9.4$\\
 004454+002509 & HELMS & 3 & $1.7 \pm 1.1$ & $7.8 \pm 3.0$ & $22.3 \pm 5.3$ & $113 \pm 23$ & $117 \pm 25$ & $116 \pm 16$\\
 004532$-$000127 & HELMS & 2 & $5.6 \pm 1.7$ & $25.0 \pm 3.9$ & $40.7 \pm 5.4$ & $107.9 \pm 8.5$ & $101.7 \pm 7.9$ & $84 \pm 11$\\
 004553$-$010038 & HELMS & 2 & $2.1 \pm 1.3$ & $8.5 \pm 3.4$ & $19.2 \pm 5.3$ & $70.7 \pm 5.6$ & $80.2 \pm 9.8$ & $87.3 \pm 9.7$\\
 004624$-$003424 & HELMS & 1 & $2.2 \pm 1.3$ & $11.1 \pm 4.3$ & $19.7 \pm 5.1$ & $33 \pm 16$ & $29 \pm 14$ & $17 \pm 13$\\
 004810+002750 & HELMS & 1 & $5.7 \pm 1.9$ & $9.5 \pm 3.9$ & $17.2 \pm 5.0$ & $48 \pm 20$ & $49 \pm 18$ & $41 \pm 12$\\
 005847$-$010025 & HELMS & 2 & $5.0 \pm 1.8$ & $10.3 \pm 4.2$ & $21.2 \pm 5.1$ & $122.8 \pm 6.2$ & $124.8 \pm 5.5$ & $72.5 \pm 4.4$\\
 005931+004243 & HELMS & 2 & $1.01 \pm 0.65$ & $7.9 \pm 3.1$ & $21.3 \pm 5.2$ & $79.3 \pm 6.3$ & $119.5 \pm 7.3$ & $139.3 \pm 4.6$\\
 010222$-$000708 & HELMS & 2 & $5.6 \pm 1.9$ & $8.3 \pm 3.2$ & $16.0 \pm 4.9$ & $95 \pm 13$ & $101.5 \pm 8.4$ & $100.9 \pm 8.1$\\
 010236+020539 & HELMS & 3 & $2.4 \pm 1.5$ & $12.3 \pm 4.9$ & $-$ & $103.9 \pm 7.0$ & $127.2 \pm 9.2$ & $86 \pm 12$\\
 010301$-$003255 & HELMS & 3 & $2.5 \pm 1.5$ & $7.5 \pm 3.1$ & $24.2 \pm 5.4$ & $160.9 \pm 4.3$ & $172.4 \pm 4.6$ & $146.1 \pm 6.5$\\
 010443$-$002209 & HELMS & 3 & $3.5 \pm 1.7$ & $9.4 \pm 3.9$ & $19.7 \pm 5.1$ & $103 \pm 16$ & $120 \pm 18$ & $120 \pm 19$\\
 010729+000114 & HERS & 2 & $2.2 \pm 1.2$ & $15.8 \pm 3.9$ & $23.0 \pm 5.0$ & $106.3 \pm 5.9$ & $113.7 \pm 5.0$ & $71.3 \pm 5.3$\\
 011640$-$000457 & HERS & 3 & $7.2 \pm 1.7$ & $21.2 \pm 3.9$ & $39.9 \pm 5.3$ & $199 \pm 20$ & $227 \pm 17$ & $169 \pm 22$\\
 011857$-$001456 & HERS & 3 & $4.4 \pm 1.8$ & $8.3 \pm 3.2$ & $16.3 \pm 4.9$ & $108.4 \pm 5.9$ & $109.8 \pm 8.3$ & $95.1 \pm 8.0$\\
 011938+001329 & HERS & 1 & $2.6 \pm 1.5$ & $8.4 \pm 3.3$ & $15.4 \pm 4.9$ & $34.4 \pm 3.7$ & $34.3 \pm 5.7$ & $15.2 \pm 5.4$\\
 012041$-$002715 & HERS & 2 & $1.31 \pm 0.80$ & $11.1 \pm 4.3$ & $40.8 \pm 5.6$ & $210.4 \pm 6.3$ & $272.7 \pm 5.5$ & $249.4 \pm 5.7$\\
 012106+003446 & HERS & 2 & $1.7 \pm 1.1$ & $9.5 \pm 4.0$ & $23.0 \pm 5.3$ & $138.6 \pm 6.7$ & $163.7 \pm 6.9$ & $122.1 \pm 7.4$\\
 013149$-$002715 & HERS & 2 & $4.4 \pm 1.8$ & $7.4 \pm 2.7$ & $16.4 \pm 5.0$ & $67 \pm 18$ & $103 \pm 20$ & $132 \pm 14$\\
 013506$-$001822 & HERS & 1 & $2.9 \pm 1.5$ & $8.4 \pm 3.3$ & $14.5 \pm 4.6$ & $33.1 \pm 3.1$ & $24.9 \pm 6.6$ & $29.9 \pm 5.9$\\
 013643$-$000655 & HERS & 3 & $5.0 \pm 1.8$ & $10.2 \pm 4.0$ & $24.5 \pm 5.1$ & $96 \pm 19$ & $110 \pm 17$ & $83 \pm 13$\\
 013655+010012 & HERS & 1 & $2.1 \pm 1.3$ & $9.3 \pm 3.5$ & $26.2 \pm 5.3$ & $44.4 \pm 9.9$ & $54.5 \pm 7.3$ & $34.4 \pm 5.9$\\
 013900+005230 & HERS & 1 & $2.3 \pm 1.3$ & $7.6 \pm 3.1$ & $23.2 \pm 5.2$ & $75.8 \pm 9.1$ & $93.8 \pm 6.3$ & $80.0 \pm 7.3$\\
 014057$-$010541 & HERS & 3 & $2.3 \pm 1.3$ & $11.1 \pm 4.2$ & $38.2 \pm 5.1$ & $176 \pm 10$ & $229 \pm 11$ & $195 \pm 13$\\
 014432$-$002446 & HERS & 2 & $2.8 \pm 1.5$ & $9.6 \pm 4.0$ & $12.9 \pm 4.5$ & $68 \pm 17$ & $61 \pm 14$ & $61 \pm 14$\\
 014530+003109 & HERS & 1 & $1.31 \pm 0.85$ & $8.8 \pm 3.5$ & $14.9 \pm 4.6$ & $43.4 \pm 9.2$ & $42.6 \pm 6.9$ & $33.1 \pm 6.1$\\
 014729+010441 & HERS & 3 & $2.1 \pm 1.3$ & $7.5 \pm 2.5$ & $19.7 \pm 4.9$ & $113 \pm 16$ & $145 \pm 17$ & $146 \pm 12$\\
 015207+013658 & HERS & 3 & $2.0 \pm 1.2$ & $10.5 \pm 4.5$ & $31 \pm 11$ & $110 \pm 17$ & $143 \pm 14$ & $141 \pm 14$\\
 015454$-$000421 & HERS & 2 & $1.6 \pm 1.0$ & $7.6 \pm 3.0$ & $24.4 \pm 5.0$ & $77.1 \pm 9.2$ & $81.6 \pm 9.4$ & $60.1 \pm 9.2$\\
 015727$-$010539 & HERS & 2 & $3.8 \pm 1.7$ & $7.8 \pm 3.1$ & $16.6 \pm 4.8$ & $83.6 \pm 9.2$ & $91 \pm 11$ & $76.6 \pm 8.7$\\
 015911$-$001349 & HERS & 4 & $1.9 \pm 1.2$ & $8.9 \pm 3.5$ & $20.3 \pm 4.9$ & $132 \pm 20$ & $151 \pm 13$ & $167 \pm 11$\\
 015943+000755 & HERS & 4 & $4.5 \pm 1.7$ & $11.1 \pm 4.2$ & $22.8 \pm 4.9$ & $155 \pm 24$ & $190 \pm 23$ & $183 \pm 34$\\
 020026+000419 & HERS & 2 & $4.7 \pm 1.8$ & $7.4 \pm 2.6$ & $16.2 \pm 4.7$ & $63.2 \pm 2.2$ & $67.4 \pm 7.0$ & $77.3 \pm 6.2$\\
 020032+011128 & HERS & 3 & $1.11 \pm 0.70$ & $9.9 \pm 3.8$ & $18.2 \pm 4.4$ & $134.1 \pm 8.1$ & $142.7 \pm 9.2$ & $131 \pm 11$\\
 020306+002807 & HERS & 2 & $1.41 \pm 0.90$ & $8.3 \pm 3.2$ & $23.4 \pm 5.0$ & $96 \pm 11$ & $93.4 \pm 7.9$ & $87.5 \pm 8.2$\\
 020325+003229 & HERS & 3 & $2.5 \pm 1.4$ & $7.6 \pm 2.7$ & $15.3 \pm 4.5$ & $118 \pm 24$ & $124 \pm 25$ & $106 \pm 22$\\
 020356+010132 & HERS & 2 & $1.9 \pm 1.2$ & $7.7 \pm 3.1$ & $18.9 \pm 4.7$ & $71 \pm 18$ & $57 \pm 24$ & $48 \pm 23$\\
 020429+010340 & HERS & 2 & $2.4 \pm 1.4$ & $9.5 \pm 3.9$ & $20.8 \pm 4.8$ & $96 \pm 16$ & $113 \pm 11$ & $97 \pm 15$\\
 020442$-$011253 & HERS & 2 & $2.8 \pm 1.4$ & $10.8 \pm 4.6$ & $21.0 \pm 4.9$ & $89.8 \pm 8.7$ & $97.8 \pm 8.8$ & $140.9 \pm 9.7$\\
 020528+000458 & HERS & 2 & $2.6 \pm 1.5$ & $7.7 \pm 2.6$ & $20.6 \pm 4.8$ & $148 \pm 10$ & $141.1 \pm 7.9$ & $127.6 \pm 9.1$\\
 020738+005845 & HERS & 4 & $1.41 \pm 0.85$ & $7.6 \pm 2.7$ & $22.2 \pm 4.8$ & $131 \pm 15$ & $142 \pm 12$ & $144 \pm 17$\\
 020935$-$005336 & HERS & 2 & $3.7 \pm 1.7$ & $7.6 \pm 2.7$ & $20.7 \pm 4.8$ & $98 \pm 12$ & $92.8 \pm 6.8$ & $98 \pm 16$\\
 020941+001557 & HERS & 6 & $17.0 \pm 1.8$ & $71.1 \pm 3.1$ & $169.1 \pm 5.3$ & $874 \pm 12$ & $1110.4 \pm 8.3$ & $1072 \pm 10$\\
 021012+003206 & HERS & 4 & $3.2 \pm 1.5$ & $11.9 \pm 4.4$ & $23.4 \pm 4.9$ & $177 \pm 15$ & $179 \pm 15$ & $141 \pm 11$\\
 021026$-$000453 & HERS & 3 & $2.5 \pm 1.4$ & $7.7 \pm 2.9$ & $15.3 \pm 4.5$ & $107 \pm 28$ & $121 \pm 26$ & $120 \pm 23$\\
 021148+004228 & HERS & 3 & $2.9 \pm 1.5$ & $8.6 \pm 3.4$ & $20.3 \pm 4.8$ & $102.3 \pm 7.4$ & $93.9 \pm 9.8$ & $128 \pm 14$\\
 021236$-$003148 & HERS & 2 & $2.2 \pm 1.3$ & $8.3 \pm 3.2$ & $20.3 \pm 4.8$ & $72 \pm 17$ & $73 \pm 13$ & $81 \pm 12$\\
 021401$-$004622 & HERS & 2 & $1.6 \pm 1.1$ & $8.3 \pm 3.2$ & $23.1 \pm 4.9$ & $159 \pm 12$ & $159.6 \pm 9.1$ & $147 \pm 11$\\
 021630$-$010411 & HERS & 3 & $1.7 \pm 1.1$ & $10.0 \pm 3.9$ & $23.1 \pm 4.9$ & $115 \pm 10$ & $126 \pm 13$ & $87 \pm 14$\\
 021645$-$002420 & HERS & 2 & $2.6 \pm 1.5$ & $7.7 \pm 2.7$ & $22.3 \pm 4.9$ & $67.9 \pm 6.3$ & $64.4 \pm 9.2$ & $48.1 \pm 8.0$\\
 021705+002041 & HERS & 3 & $4.2 \pm 1.7$ & $9.3 \pm 3.6$ & $18.8 \pm 4.7$ & $103 \pm 20$ & $99 \pm 15$ & $75 \pm 17$\\
 021715+005618 & HERS & 2 & $7.2 \pm 1.8$ & $11.8 \pm 4.3$ & $22.8 \pm 4.8$ & $81 \pm 12$ & $84 \pm 10$ & $44.5 \pm 9.6$\\
 021921$-$010747 & HERS & 2 & $3.2 \pm 1.6$ & $7.6 \pm 2.8$ & $15.2 \pm 4.5$ & $65.4 \pm 3.6$ & $52.3 \pm 7.5$ & $65.7 \pm 8.2$\\
 232748$-$012045 & HELMS & 1 & $6.3 \pm 2.1$ & $10.7 \pm 4.1$ & $19.9 \pm 5.6$ & $34.4 \pm 6.3$ & $58.1 \pm 7.6$ & $56.9 \pm 7.4$\\
 232916$-$003755 & HELMS & 1 & $3.6 \pm 1.9$ & $8.6 \pm 3.4$ & $19.7 \pm 6.1$ & $37.5 \pm 5.9$ & $33.3 \pm 6.4$ & $24.3 \pm 8.0$\\
 233802$-$011904 & HELMS & 2 & $4.0 \pm 1.9$ & $14.1 \pm 5.4$ & $21.9 \pm 5.7$ & $102 \pm 21$ & $75 \pm 17$ & $39 \pm 12$\\
 234648$-$000519 & HERS & 3 & $4.9 \pm 2.2$ & $10.0 \pm 3.8$ & $16.7 \pm 5.3$ & $141 \pm 16$ & $144 \pm 13$ & $137 \pm 13$\\
\end{longtable}

\begin{longtable}{c c c c c c}
 \caption{ACT-S ID, ACT S$_{1400}$ (column 2), SMA flux densities (columns 3-6) } \\
 \hline
 ACT-S ID & $S_{\mathrm{1400}}$ & $S_{\mathrm{1332}}$ & $S_{\mathrm{1329}}$ & $S_{\mathrm{1312}}$ & $S_{\mathrm{1110}}$ \\ (J2000) & (mJy) & (mJy) & (mJy) & (mJy) & (mJy)\\ \hline\endfirsthead
 \caption{(Continued)}\\
 \hline
 ACT-S ID & $S_{\mathrm{1400}}$ & $S_{\mathrm{1329}}$ & $S_{\mathrm{1312}}$ & $S_{\mathrm{1110}}$ \\ (J2000) & (mJy)& (mJy) & (mJy) & (mJy) & (mJy))\\ 
 \hline
 \endhead
 \label{tab:smacatalog}
001133$-$001835 & $20.0 \pm 3.5$ & $-$ & $-$ & $15.5 \pm 3.5$ & $-$\\
 002147+002445 & $8.4 \pm 3.3$ & $-$ & $8.00 \pm 0.89$ & $-$ & $-$\\
 002220$-$015523 & $26.3 \pm 3.8$ & $-$ & $-$ & $-$ & $26.9 \pm 2.2$\\
 002318$-$001554 & $11.1 \pm 4.3$ & $-$ & $6.10 \pm 0.40$ & $-$ & $-$\\
 003337+000353 & $8.8 \pm 3.5$ & $-$ & $16.0 \pm 3.1$ & $-$ & $-$\\
 003814$-$002255 & $23.5 \pm 3.3$ & $17.1 \pm 1.0$ & $-$ & $-$ & $-$\\
 003929+002422 & $17.7 \pm 4.2$ & $-$ & $-$ & $23.1 \pm 3.1$ & $-$\\
 004410+011818 & $34.7 \pm 4.2$ & $-$ & $-$ & $-$ & $47.3 \pm 4.4$\\
 004532$-$000127 & $25.0 \pm 3.9$ & $-$ & $-$ & $27.5 \pm 2.9$ & $-$\\
 005847$-$010025 & $10.3 \pm 4.2$ & $-$ & $14.2 \pm 2.6$ & $-$ & $-$\\
 010729+000114 & $15.8 \pm 3.9$ & $-$ & $-$ & $6.60 \pm 0.96$ & $-$\\
 011640$-$000457 & $21.2 \pm 3.9$ & $-$ & $-$ & $18.8 \pm 4.0$ & $-$\\
 011938+001329 & $8.4 \pm 3.3$ & $-$ & $4.20 \pm 0.40$ & $-$ & $-$\\
 013506$-$001822 & $8.4 \pm 3.3$ & $-$ & $5.10 \pm 0.65$ & $-$ & $-$\\
 014432$-$002446 & $9.6 \pm 4.0$ & $-$ & $3.50 \pm 0.63$ & $-$ & $-$\\
 014530+003109 & $8.8 \pm 3.5$ & $-$ & $3.40 \pm 0.43$ & $-$ & $-$\\
 020941+001557 & $71.1 \pm 3.1$ & $-$ & $-$ & $29.0 \pm 4.3$ & $-$\\
 021236$-$003148 & $8.3 \pm 3.2$ & $-$ & $4.40 \pm 0.64$ & $-$ & $-$\\
 021715+005618 & $11.8 \pm 4.3$ & $-$ & $8.20 \pm 0.90$ & $-$ & $-$\\
\end{longtable}

\begin{longtable}{c c c c c c c c}
\caption{Median, 16th and 84th percentile values for the fit parameters used to model the ensemble SEDs}\\
 \hline
 ACT-S ID & $z_{\mathrm{phot}}$ & $\sqrt{\mu} d$ & $\log_{10} (\mu M_\mathrm{d}/M_\odot)$ & $T_\mathrm{c}$ & $\log_{10} (\mu L_\mathrm{IR}/L_\odot)$ & $\tau_{100}$ & $\chi^2$\\(J2000) & & (kpc) & & (K) & & & \\ \hline\endfirsthead
\caption{(Continued)}\\
 \hline
 ACT-S ID & $z_{\mathrm{phot}}$ & $\sqrt{\mu} d$ & $\log_{10} (\mu M_\mathrm{d}/M_\odot)$ & $T_\mathrm{c}$ & $\log_{10} (\mu L_\mathrm{IR}/L_\odot)$ & $\tau_{100}$ & $\chi^2$\\(J2000) & & (kpc) & & (K) & & & \\ \hline\endhead
 \label{tab:fitparams}
 001133$-$001835 & $3.1^{+1.2}_{-0.9}$ & $2.8^{+0.9}_{-0.5}$ & $9.34^{+0.27}_{-0.3}$ & $43.4^{+10.9}_{-8.0}$ & $13.65^{+0.31}_{-0.33}$ & $8.0^{+7.5}_{-4.7}$ & 9.83\\
 001855$-$010649 & $4.1^{+1.3}_{-1.1}$ & $5.9^{+2.9}_{-2.3}$ & $9.1^{+0.28}_{-0.25}$ & $34.0^{+7.8}_{-5.8}$ & $13.61^{+0.33}_{-0.31}$ & $1.0^{+2.4}_{-0.6}$ & 2.48\\
 002147+002445 & $4.3^{+1.5}_{-1.2}$ & $6.1^{+3.2}_{-2.7}$ & $9.11^{+0.26}_{-0.27}$ & $33.8^{+7.2}_{-5.9}$ & $13.57^{+0.3}_{-0.32}$ & $1.0^{+3.0}_{-0.6}$ & 2.04\\
 002220$-$015523 & $4.4^{+1.5}_{-1.0}$ & $4.8^{+3.6}_{-0.9}$ & $9.29^{+0.26}_{-0.28}$ & $40.8^{+8.6}_{-6.9}$ & $13.91^{+0.29}_{-0.27}$ & $2.5^{+2.2}_{-1.9}$ & 5.58\\
 002220$-$015523$^\dag$ & $5.16$ & $4.7^{+4.7}_{-1.0}$ & $9.15 \pm 0.08$ & $46.0^{+5.7}_{-10.3}$ & $14.06 \pm 0.03$ & $1.7^{+1.0}_{-1.1}$ & 3.44\\
 002318$-$001554 & $3.9^{+1.3}_{-1.1}$ & $6.1^{+3.4}_{-2.7}$ & $9.12^{+0.24}_{-0.26}$ & $35.2^{+7.5}_{-6.1}$ & $13.66^{+0.29}_{-0.32}$ & $1.0^{+2.9}_{-0.7}$ & 10.53\\
 002741$-$011651 & $4.1^{+1.4}_{-1.1}$ & $5.8^{+3.4}_{-2.3}$ & $9.09^{+0.25}_{-0.26}$ & $32.0^{+6.8}_{-5.5}$ & $13.46^{+0.3}_{-0.32}$ & $1.1^{+2.3}_{-0.7}$ & 4.56\\
 003337+000353 & $3.5^{+1.4}_{-1.0}$ & $4.5^{+4.6}_{-1.8}$ & $9.11 \pm 0.26$ & $33.8^{+7.8}_{-5.9}$ & $13.44^{+0.33}_{-0.35}$ & $1.7^{+5.7}_{-1.4}$ & 4.38\\
 003648$-$002052 & $2.3^{+1.2}_{-0.8}$ & $3.0^{+5.9}_{-0.8}$ & $9.2^{+0.22}_{-0.26}$ & $36.1^{+9.0}_{-6.5}$ & $13.45^{+0.4}_{-0.43}$ & $4.7^{+7.7}_{-4.3}$ & 22.13\\
 003757$-$010621 & $3.3^{+1.3}_{-1.0}$ & $5.6^{+3.9}_{-2.1}$ & $9.14 \pm 0.26$ & $36.9^{+8.0}_{-6.6}$ & $13.75^{+0.32}_{-0.35}$ & $1.2^{+3.2}_{-0.9}$ & 7.28\\
 003814$-$002255 & $3.8^{+1.2}_{-1.0}$ & $3.3^{+1.0}_{-0.6}$ & $9.38^{+0.26}_{-0.27}$ & $45.5^{+9.9}_{-9.3}$ & $13.81^{+0.26}_{-0.29}$ & $6.3^{+5.8}_{-3.5}$ & 1.34\\
 003929+002422 & $3.2^{+1.2}_{-0.9}$ & $3.8^{+1.5}_{-0.5}$ & $9.32^{+0.25}_{-0.29}$ & $42.2^{+9.2}_{-7.9}$ & $13.8^{+0.32}_{-0.33}$ & $4.1^{+3.4}_{-2.6}$ & 12.13\\
 003943$-$003952 & $3.7^{+1.3}_{-1.1}$ & $5.5^{+3.9}_{-2.5}$ & $9.13 \pm 0.25$ & $34.4^{+7.8}_{-6.4}$ & $13.56^{+0.31}_{-0.35}$ & $1.3^{+4.2}_{-0.9}$ & 5.08\\
 004033+000228 & $3.7^{+1.5}_{-1.3}$ & $4.5^{+4.4}_{-2.4}$ & $9.12^{+0.28}_{-0.27}$ & $33.9^{+8.3}_{-6.4}$ & $13.39^{+0.33}_{-0.39}$ & $1.8^{+10.3}_{-1.4}$ & 2.34\\
 004338$-$005320 & $3.2^{+1.3}_{-1.0}$ & $4.6^{+4.9}_{-2.1}$ & $9.11^{+0.25}_{-0.27}$ & $34.1^{+8.7}_{-6.7}$ & $13.45^{+0.34}_{-0.36}$ & $1.6^{+6.7}_{-1.3}$ & 8.25\\
 004410+011818 & $4.6^{+1.3}_{-1.2}$ & $4.9^{+1.0}_{-0.6}$ & $9.45^{+0.29}_{-0.28}$ & $46.0^{+9.7}_{-9.3}$ & $14.14^{+0.25}_{-0.29}$ & $3.3^{+2.7}_{-1.5}$ & 4.03\\
 004410+011818$^\dag$ & $4.16$ & $5.0^{+0.8}_{-0.6}$ & $9.55^{+0.05}_{-0.06}$ & $43.5 \pm 3.3$ & $14.04 \pm 0.03$ & $4.0^{+1.2}_{-1.0}$ & 0.55\\
 004454+002509 & $3.0^{+1.2}_{-1.0}$ & $5.3^{+3.8}_{-2.5}$ & $9.12^{+0.24}_{-0.25}$ & $34.9^{+8.4}_{-6.4}$ & $13.58^{+0.34}_{-0.4}$ & $1.3^{+4.9}_{-1.0}$ & 3.8\\
 004532$-$000127 & $4.0^{+1.2}_{-0.9}$ & $3.4^{+0.8}_{-0.5}$ & $9.37^{+0.25}_{-0.27}$ & $44.7^{+9.8}_{-8.0}$ & $13.77^{+0.25}_{-0.27}$ & $6.0^{+4.7}_{-2.9}$ & 2.37\\
 004553$-$010038 & $2.3^{+1.1}_{-0.8}$ & $2.2^{+2.0}_{-0.4}$ & $9.21^{+0.24}_{-0.29}$ & $38.0^{+10.0}_{-8.0}$ & $13.22^{+0.39}_{-0.41}$ & $9.4^{+11.1}_{-7.8}$ & 7.67\\
 004624$-$003424 & $4.9^{+1.9}_{-1.5}$ & $5.0^{+4.0}_{-2.5}$ & $9.1 \pm 0.28$ & $32.0^{+7.3}_{-5.8}$ & $13.3^{+0.32}_{-0.34}$ & $1.5^{+6.7}_{-1.1}$ & 1.23\\
 004810+002750 & $3.9^{+1.6}_{-1.2}$ & $2.4^{+5.6}_{-0.9}$ & $9.26^{+0.33}_{-0.32}$ & $39.1^{+14.7}_{-8.7}$ & $13.41^{+0.32}_{-0.36}$ & $8.6^{+29.7}_{-8.1}$ & 5.21\\
 005847$-$010025 & $4.3^{+1.2}_{-1.0}$ & $6.8^{+2.3}_{-1.8}$ & $9.16^{+0.26}_{-0.25}$ & $35.7^{+7.2}_{-6.2}$ & $13.83^{+0.25}_{-0.29}$ & $0.9^{+1.0}_{-0.5}$ & 2.4\\
 005931+004243 & $1.9^{+1.1}_{-0.8}$ & $2.4^{+6.8}_{-0.7}$ & $9.08^{+0.19}_{-0.23}$ & $35.6^{+10.3}_{-7.5}$ & $13.28^{+0.41}_{-0.51}$ & $5.5^{+12.2}_{-5.1}$ & 48.61\\
 010222$-$000708 & $2.5^{+1.3}_{-0.8}$ & $2.7^{+4.6}_{-0.7}$ & $9.18^{+0.26}_{-0.29}$ & $38.0^{+9.7}_{-7.1}$ & $13.4^{+0.38}_{-0.4}$ & $5.9^{+8.9}_{-5.3}$ & 13.12\\
 010236+020539 & $3.7^{+1.3}_{-1.0}$ & $6.4^{+3.0}_{-2.6}$ & $9.12^{+0.24}_{-0.25}$ & $35.2^{+7.1}_{-6.0}$ & $13.71^{+0.31}_{-0.33}$ & $0.9^{+2.3}_{-0.6}$ & 6.73\\
 010301$-$003255 & $3.4^{+1.1}_{-0.9}$ & $5.9^{+3.1}_{-1.6}$ & $9.19^{+0.24}_{-0.25}$ & $37.3^{+8.1}_{-7.1}$ & $13.84^{+0.28}_{-0.31}$ & $1.2^{+1.7}_{-0.7}$ & 6.2\\
 010443$-$002209 & $3.0^{+1.3}_{-1.0}$ & $5.1^{+4.4}_{-2.6}$ & $9.15^{+0.24}_{-0.26}$ & $35.6^{+8.8}_{-7.0}$ & $13.59^{+0.35}_{-0.4}$ & $1.5^{+6.7}_{-1.1}$ & 6.95\\
 010729+000114 & $4.0^{+1.3}_{-1.0}$ & $6.7^{+2.8}_{-2.2}$ & $9.14 \pm 0.24$ & $34.9^{+7.1}_{-5.9}$ & $13.74^{+0.28}_{-0.29}$ & $0.9^{+1.4}_{-0.6}$ & 6.34\\
 010729+000114$^\dag$ & $3.33$ & $5.9^{+2.9}_{-1.7}$ & $9.3^{+0.1}_{-0.09}$ & $31.6^{+4.7}_{-4.1}$ & $13.55 \pm 0.03$ & $1.5^{+1.2}_{-0.6}$ & 2.06\\
 011640$-$000457 & $4.1^{+1.4}_{-1.1}$ & $6.7^{+3.1}_{-2.3}$ & $9.26^{+0.25}_{-0.27}$ & $41.5^{+8.6}_{-7.3}$ & $14.11 \pm 0.29$ & $1.2^{+2.3}_{-0.8}$ & 4.47\\
 011640$-$000457$^\dag$ & $3.79$ & $6.4^{+3.0}_{-2.1}$ & $9.34^{+0.09}_{-0.1}$ & $39.0^{+8.2}_{-5.3}$ & $14.02^{+0.05}_{-0.04}$ & $1.5^{+1.4}_{-0.7}$ & 1.71\\
 011857$-$001456 & $3.5^{+1.3}_{-1.1}$ & $6.5^{+3.3}_{-2.8}$ & $9.11^{+0.26}_{-0.25}$ & $34.8^{+7.1}_{-6.2}$ & $13.68^{+0.32}_{-0.35}$ & $0.9^{+2.6}_{-0.6}$ & 16.08\\
 011938+001329 & $4.3^{+1.6}_{-1.2}$ & $3.8^{+5.0}_{-1.5}$ & $9.07^{+0.28}_{-0.26}$ & $32.3^{+7.8}_{-6.0}$ & $13.24^{+0.32}_{-0.31}$ & $2.3^{+6.9}_{-2.0}$ & 2.4\\
 012041$-$002715 & $3.1^{+1.0}_{-0.8}$ & $6.1^{+2.5}_{-1.6}$ & $9.24 \pm 0.24$ & $39.1^{+8.2}_{-7.5}$ & $13.96^{+0.29}_{-0.32}$ & $1.3^{+1.6}_{-0.7}$ & 11.24\\
 012106+003446 & $3.5^{+1.1}_{-0.9}$ & $6.5^{+2.7}_{-2.0}$ & $9.17^{+0.23}_{-0.25}$ & $35.6^{+7.2}_{-6.2}$ & $13.79^{+0.28}_{-0.3}$ & $1.0^{+1.4}_{-0.6}$ & 3.92\\
 013149$-$002715 & $1.4^{+0.8}_{-0.6}$ & $1.5^{+1.4}_{-0.5}$ & $9.31^{+0.22}_{-0.25}$ & $42.6^{+13.0}_{-10.5}$ & $13.09^{+0.39}_{-0.48}$ & $26.8^{+54.3}_{-22.4}$ & 5.14\\
 013506$-$001822 & $3.3^{+1.3}_{-1.0}$ & $1.8^{+0.7}_{-0.4}$ & $9.28^{+0.26}_{-0.3}$ & $39.6^{+11.4}_{-8.6}$ & $13.07^{+0.32}_{-0.35}$ & $16.2^{+19.8}_{-9.8}$ & 3.98\\
 013643$-$000655 & $3.7^{+1.4}_{-1.1}$ & $5.5^{+3.8}_{-2.5}$ & $9.17 \pm 0.27$ & $36.5^{+8.3}_{-6.9}$ & $13.67^{+0.33}_{-0.34}$ & $1.4^{+5.0}_{-1.0}$ & 5.67\\
 013655+010012 & $3.7^{+1.6}_{-1.1}$ & $3.4^{+5.2}_{-1.2}$ & $9.11^{+0.27}_{-0.28}$ & $34.9^{+8.2}_{-6.7}$ & $13.35^{+0.35}_{-0.34}$ & $3.2^{+7.4}_{-2.8}$ & 5.11\\
 013900+005230 & $3.0^{+1.3}_{-1.0}$ & $4.4^{+5.0}_{-2.0}$ & $9.12^{+0.25}_{-0.27}$ & $34.4^{+8.2}_{-6.6}$ & $13.46^{+0.34}_{-0.4}$ & $1.8^{+7.4}_{-1.5}$ & 6.23\\
 014057$-$010541 & $3.4^{+1.1}_{-1.0}$ & $6.7^{+3.0}_{-2.5}$ & $9.22^{+0.23}_{-0.25}$ & $37.8^{+8.3}_{-6.5}$ & $13.94^{+0.28}_{-0.33}$ & $1.1^{+2.0}_{-0.7}$ & 4.93\\
 014432$-$002446 & $3.2^{+1.5}_{-1.2}$ & $3.8^{+5.1}_{-2.0}$ & $9.1 \pm 0.27$ & $34.3^{+9.6}_{-6.6}$ & $13.35^{+0.36}_{-0.43}$ & $2.4^{+12.5}_{-2.0}$ & 2.45\\
 014530+003109 & $3.4^{+1.5}_{-1.1}$ & $3.7^{+5.0}_{-1.7}$ & $9.04 \pm 0.26$ & $32.3^{+7.9}_{-6.0}$ & $13.2^{+0.36}_{-0.39}$ & $2.3^{+8.5}_{-2.0}$ & 2.68\\
 014729+010441 & $2.7^{+1.2}_{-1.0}$ & $5.4^{+3.9}_{-2.7}$ & $9.11^{+0.24}_{-0.25}$ & $35.5^{+7.6}_{-6.4}$ & $13.6^{+0.37}_{-0.42}$ & $1.2^{+5.7}_{-0.9}$ & 11.16\\
 015207+013658 & $2.8^{+1.3}_{-1.0}$ & $4.7^{+4.5}_{-2.2}$ & $9.15^{+0.24}_{-0.26}$ & $36.4^{+9.0}_{-7.0}$ & $13.6^{+0.37}_{-0.4}$ & $1.8^{+7.1}_{-1.4}$ & 6.66\\
 015454$-$000421 & $3.5^{+1.3}_{-1.0}$ & $5.3^{+3.8}_{-2.2}$ & $9.09^{+0.24}_{-0.25}$ & $33.5^{+7.0}_{-5.8}$ & $13.49^{+0.31}_{-0.35}$ & $1.3^{+3.5}_{-1.0}$ & 4.55\\
 015727$-$010539 & $3.3^{+1.2}_{-1.0}$ & $4.6^{+4.5}_{-2.0}$ & $9.1 \pm 0.25$ & $34.7^{+7.8}_{-6.1}$ & $13.51^{+0.32}_{-0.37}$ & $1.6^{+5.0}_{-1.3}$ & 2.93\\
 015911$-$001349 & $2.4^{+1.3}_{-0.9}$ & $3.6^{+4.8}_{-1.3}$ & $9.18^{+0.23}_{-0.28}$ & $36.9^{+9.5}_{-7.5}$ & $13.53^{+0.41}_{-0.46}$ & $3.1^{+7.8}_{-2.7}$ & 4.86\\
 015943+000755 & $3.1^{+1.3}_{-1.1}$ & $6.1^{+3.5}_{-3.0}$ & $9.2^{+0.25}_{-0.26}$ & $37.8^{+8.5}_{-7.0}$ & $13.81^{+0.34}_{-0.4}$ & $1.2^{+4.6}_{-0.9}$ & 6.64\\
 020026+000419 & $2.2^{+0.8}_{-0.7}$ & $1.9^{+0.3}_{-0.2}$ & $9.28 \pm 0.24$ & $40.6^{+10.5}_{-8.7}$ & $13.14^{+0.32}_{-0.4}$ & $14.7^{+13.4}_{-6.5}$ & 5.67\\
 020032+011128 & $2.9^{+1.0}_{-0.8}$ & $5.4^{+3.3}_{-1.8}$ & $9.17^{+0.23}_{-0.24}$ & $34.5^{+7.4}_{-6.3}$ & $13.63^{+0.31}_{-0.34}$ & $1.4^{+2.3}_{-0.9}$ & 7.02\\
 020306+002807 & $3.1^{+1.2}_{-0.9}$ & $3.9^{+4.6}_{-1.3}$ & $9.11^{+0.24}_{-0.25}$ & $35.4^{+8.3}_{-6.4}$ & $13.5^{+0.33}_{-0.36}$ & $2.4^{+4.5}_{-2.0}$ & 5.69\\
 020325+003229 & $3.1^{+1.2}_{-1.0}$ & $6.1^{+3.2}_{-2.7}$ & $9.11^{+0.23}_{-0.25}$ & $33.8^{+7.4}_{-5.9}$ & $13.6^{+0.32}_{-0.38}$ & $1.0^{+2.9}_{-0.6}$ & 1.88\\
 020356+010132 & $3.8^{+1.5}_{-1.2}$ & $5.3^{+3.9}_{-2.6}$ & $9.08^{+0.26}_{-0.25}$ & $33.5^{+7.5}_{-6.0}$ & $13.44^{+0.32}_{-0.37}$ & $1.2^{+5.4}_{-0.9}$ & 1.52\\
 020429+010340 & $3.3^{+1.3}_{-1.1}$ & $5.8^{+3.6}_{-2.8}$ & $9.13^{+0.26}_{-0.28}$ & $34.6^{+7.7}_{-6.4}$ & $13.6^{+0.35}_{-0.41}$ & $1.2^{+4.4}_{-0.8}$ & 6.9\\
 020442$-$011253 & $1.6^{+0.7}_{-0.6}$ & $1.7 \pm 0.3$ & $9.38^{+0.19}_{-0.22}$ & $43.8^{+11.8}_{-9.1}$ & $13.17^{+0.38}_{-0.48}$ & $23.1^{+25.0}_{-11.1}$ & 5.21\\
 020528+000458 & $3.2^{+1.2}_{-0.9}$ & $5.4^{+3.4}_{-1.6}$ & $9.17^{+0.24}_{-0.25}$ & $36.0^{+7.5}_{-6.2}$ & $13.71^{+0.31}_{-0.34}$ & $1.4^{+2.3}_{-1.0}$ & 6.38\\
 020738+005845 & $2.9^{+1.1}_{-0.9}$ & $5.6^{+3.8}_{-2.5}$ & $9.12^{+0.22}_{-0.24}$ & $35.5^{+7.8}_{-6.3}$ & $13.65^{+0.32}_{-0.37}$ & $1.2^{+3.8}_{-0.8}$ & 8.5\\
 020935$-$005336 & $3.1^{+1.4}_{-1.0}$ & $3.7^{+4.8}_{-1.3}$ & $9.14 \pm 0.26$ & $36.2^{+9.1}_{-6.8}$ & $13.5^{+0.34}_{-0.39}$ & $3.0^{+6.9}_{-2.6}$ & 5.23\\
 020941+001557 & $4.0^{+1.1}_{-0.9}$ & $7.5^{+1.4}_{-1.1}$ & $9.59^{+0.25}_{-0.26}$ & $58.2^{+12.0}_{-10.1}$ & $14.91^{+0.24}_{-0.26}$ & $2.0^{+1.3}_{-0.9}$ & 2.04\\
 020941+001557$^\dag$ & $2.55$ & $8.1^{+1.2}_{-0.9}$ & $10.0 \pm 0.04$ & $43.2^{+3.4}_{-3.3}$ & $14.45 \pm 0.04$ & $4.4^{+0.9}_{-0.8}$ & 1.26\\
 021012+003206 & $3.7^{+1.2}_{-1.0}$ & $6.8^{+2.6}_{-2.0}$ & $9.19^{+0.25}_{-0.26}$ & $37.2^{+7.9}_{-6.5}$ & $13.92^{+0.29}_{-0.33}$ & $1.0^{+1.3}_{-0.6}$ & 3.05\\
 021026$-$000453 & $2.8^{+1.2}_{-1.1}$ & $5.2^{+4.1}_{-2.8}$ & $9.12^{+0.23}_{-0.25}$ & $34.6^{+8.1}_{-6.4}$ & $13.52^{+0.35}_{-0.45}$ & $1.4^{+7.0}_{-1.0}$ & 3.33\\
 021148+004228 & $2.3^{+1.1}_{-0.7}$ & $2.6^{+1.1}_{-0.5}$ & $9.27^{+0.22}_{-0.26}$ & $38.6^{+9.0}_{-7.1}$ & $13.33^{+0.36}_{-0.37}$ & $7.9^{+8.8}_{-5.1}$ & 9.4\\
 021236$-$003148 & $2.7^{+1.3}_{-1.0}$ & $2.8^{+5.9}_{-1.0}$ & $9.18^{+0.25}_{-0.28}$ & $36.1^{+9.4}_{-7.3}$ & $13.32^{+0.37}_{-0.41}$ & $5.6^{+15.7}_{-5.1}$ & 4.84\\
 021401$-$004622 & $3.1^{+1.1}_{-0.9}$ & $5.6^{+3.2}_{-1.8}$ & $9.17^{+0.24}_{-0.25}$ & $36.2^{+7.5}_{-6.4}$ & $13.74^{+0.31}_{-0.34}$ & $1.4^{+2.2}_{-0.9}$ & 6.34\\
 021630$-$010411 & $3.6^{+1.2}_{-1.0}$ & $6.0^{+3.0}_{-2.0}$ & $9.13^{+0.24}_{-0.25}$ & $35.1^{+7.2}_{-5.9}$ & $13.7^{+0.29}_{-0.31}$ & $1.1^{+1.8}_{-0.7}$ & 1.87\\
 021645$-$002420 & $3.8^{+1.4}_{-1.1}$ & $5.6^{+3.8}_{-2.7}$ & $9.09^{+0.25}_{-0.26}$ & $33.3^{+7.0}_{-5.8}$ & $13.48^{+0.32}_{-0.34}$ & $1.1^{+4.0}_{-0.8}$ & 10.07\\
 021705+002041 & $3.8^{+1.3}_{-1.1}$ & $6.3^{+3.2}_{-2.9}$ & $9.12^{+0.25}_{-0.26}$ & $34.7^{+7.3}_{-6.0}$ & $13.65^{+0.31}_{-0.35}$ & $1.0^{+3.0}_{-0.6}$ & 3.96\\
 021715+005618 & $4.5^{+1.5}_{-1.2}$ & $6.2^{+3.4}_{-2.9}$ & $9.15^{+0.26}_{-0.25}$ & $36.1^{+7.8}_{-6.2}$ & $13.7^{+0.29}_{-0.3}$ & $1.1^{+3.8}_{-0.7}$ & 11.39\\
 021921$-$010747 & $2.8^{+1.3}_{-0.9}$ & $2.6^{+4.1}_{-0.5}$ & $9.15^{+0.25}_{-0.27}$ & $35.9^{+8.1}_{-6.8}$ & $13.25^{+0.36}_{-0.35}$ & $5.9^{+7.0}_{-5.3}$ & 13.45\\
 232748$-$012045 & $2.7^{+1.2}_{-1.0}$ & $1.4^{+0.3}_{-0.2}$ & $9.53^{+0.26}_{-0.27}$ & $50.9^{+14.7}_{-11.7}$ & $13.26^{+0.36}_{-0.43}$ & $50.2^{+50.7}_{-24.5}$ & 3.92\\
 232916$-$003755 & $4.1^{+1.8}_{-1.2}$ & $2.9^{+5.5}_{-1.0}$ & $9.15 \pm 0.3$ & $35.3^{+9.4}_{-7.3}$ & $13.27 \pm 0.33$ & $4.7^{+13.5}_{-4.3}$ & 2.22\\
 233802$-$011904 & $4.7^{+1.5}_{-1.3}$ & $6.5^{+3.0}_{-2.4}$ & $9.14 \pm 0.26$ & $35.1^{+6.8}_{-6.0}$ & $13.71^{+0.28}_{-0.32}$ & $1.0^{+2.1}_{-0.6}$ & 1.13\\
 234648$-$000519 & $3.2^{+1.2}_{-1.1}$ & $6.0^{+3.6}_{-2.6}$ & $9.15 \pm 0.25$ & $35.7^{+8.2}_{-6.3}$ & $13.73^{+0.32}_{-0.39}$ & $1.1^{+3.2}_{-0.8}$ & 5.91\\
\end{longtable}
\tablenotetext{\dag}{Denotes a fit using fixed spectroscopic redshift}

\section{Full SMA cross matched results}
\label{appendixB}
Below we provide the SEDs of nineteen sources with SMA observations, as well as thumbnail cutouts from ACT, \herschel{}, and several near-infrared and optical telescopes where valid, including Spitzer Space Telescope, the VISTA Hemisphere Survey, and Pan-STARRS. 
The SEDs shown for the five sources with known spectroscopic redshifts, we fixed our redshifts to these values when performing the SED fits. 
For other sources, we allowed the redshift to vary and obtained the photometric redshift as a fit parameter.

Black points are from ACT, blue points are the \herschel{} ensemble fluxes, blue curves are the ``median" SEDs generated from the median parameter values of the posterior distributions, magenta triangles are the flux densities of the brightest PCAT counterpart, and green stars are the flux densities of the SMA source that lies closest to the ACT source center. 
These SMA source locations are plotted as green stars on the ACT and \herschel{} thumbnails, and the SMA contours of this source are overplotted on the near-IR and optical thumbnails. The contours correspond to 3, 4, 5 and 6 times the RMS of each map; we also include a contour for double the RMS in the case of J014530+003109. The SMA clean beam is shown in the bottom right corner as an ellipse. If the source has a VLA detection, its location is indicated with a red pentagon.


\begin{figure*}
\includegraphics[width=\columnwidth]{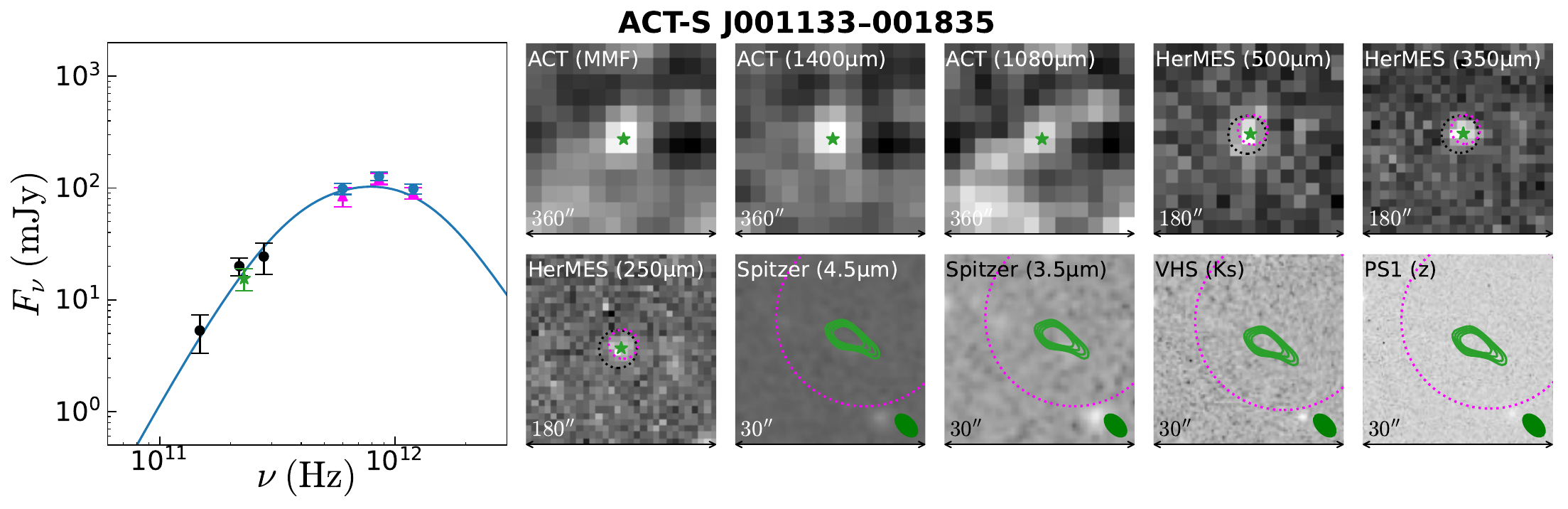}
\includegraphics[width=\columnwidth]{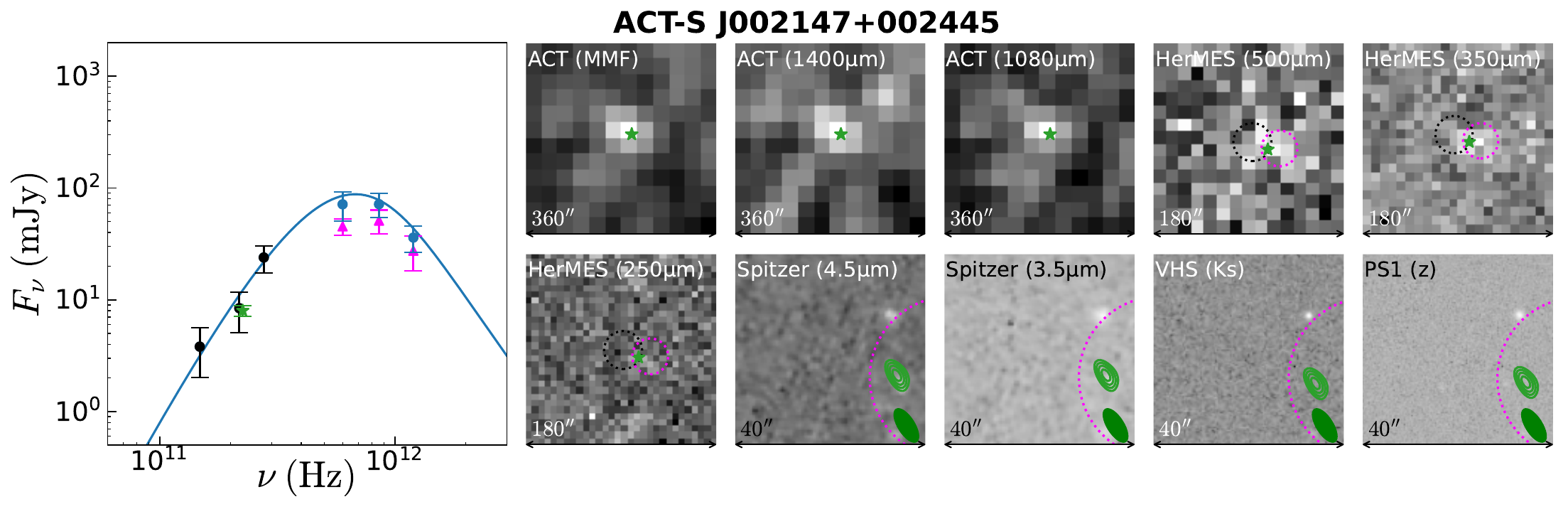}
\includegraphics[width=\columnwidth]{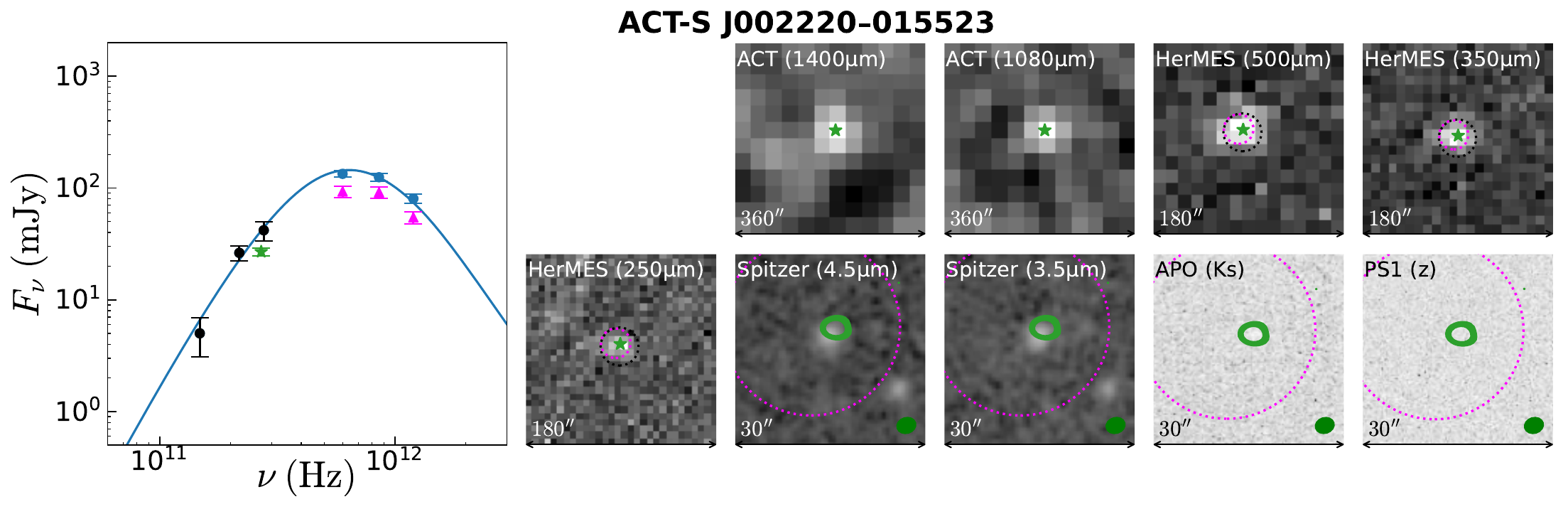}
\caption{ACT-selected DSFGs observed with the SMA. Left: SED using the ACT (black), \herschel{} ensemble (blue), SMA (green), and brightest PCAT-predicted counterpart (magenta) flux densities, and the best-fitting model SED (blue curve). Right: ACT and \herschel{} thumbnails, as well as near-IR and optical cutouts from Spitzer, Apache Point Observatory 1.5~m telescope, VHA, and PS1, as labeled. Green star is the location of the SMA source detected nearest to the ACT center location. Black dashed curve on the \herschel{} thumbnails is the 30" radius used to quantify PCAT counterparts, and the pink magenta circles on the \herschel{}, near-IR and optical images are the 90\% flux radius of the PCAT posterior images, centered on the location of the brightest counterpart.}
\label{fig:thumbnaila}
\end{figure*}

\begin{figure*}
\ContinuedFloat
\includegraphics[width=\columnwidth]{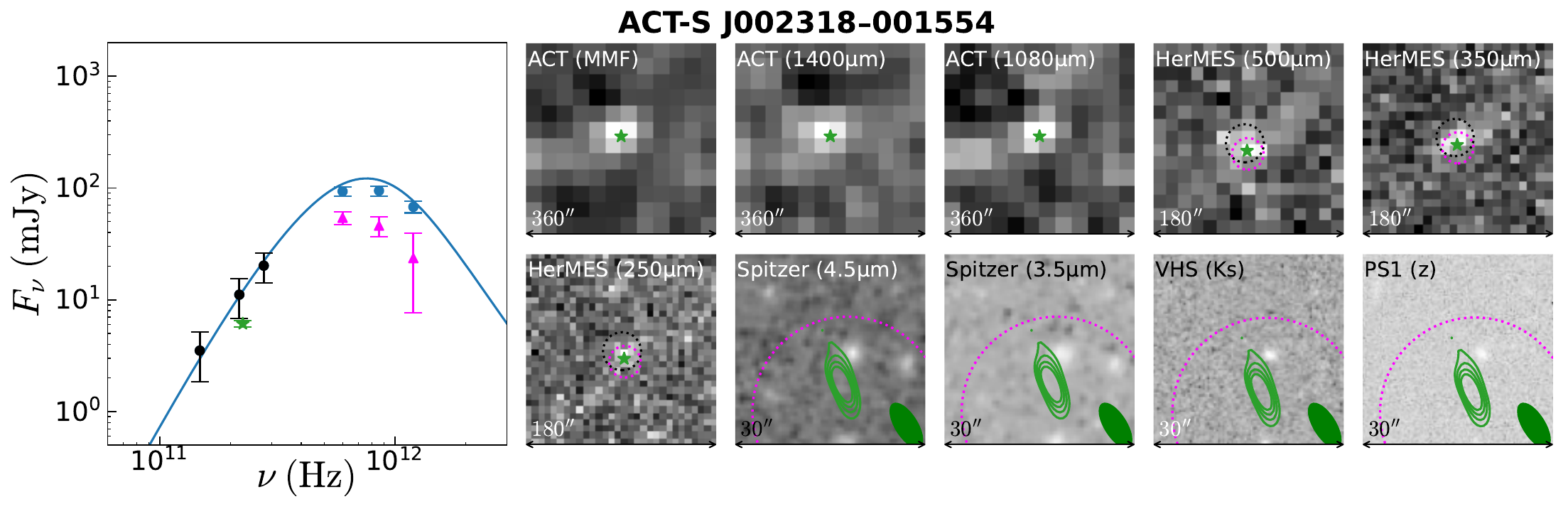}
\includegraphics[width=\columnwidth]{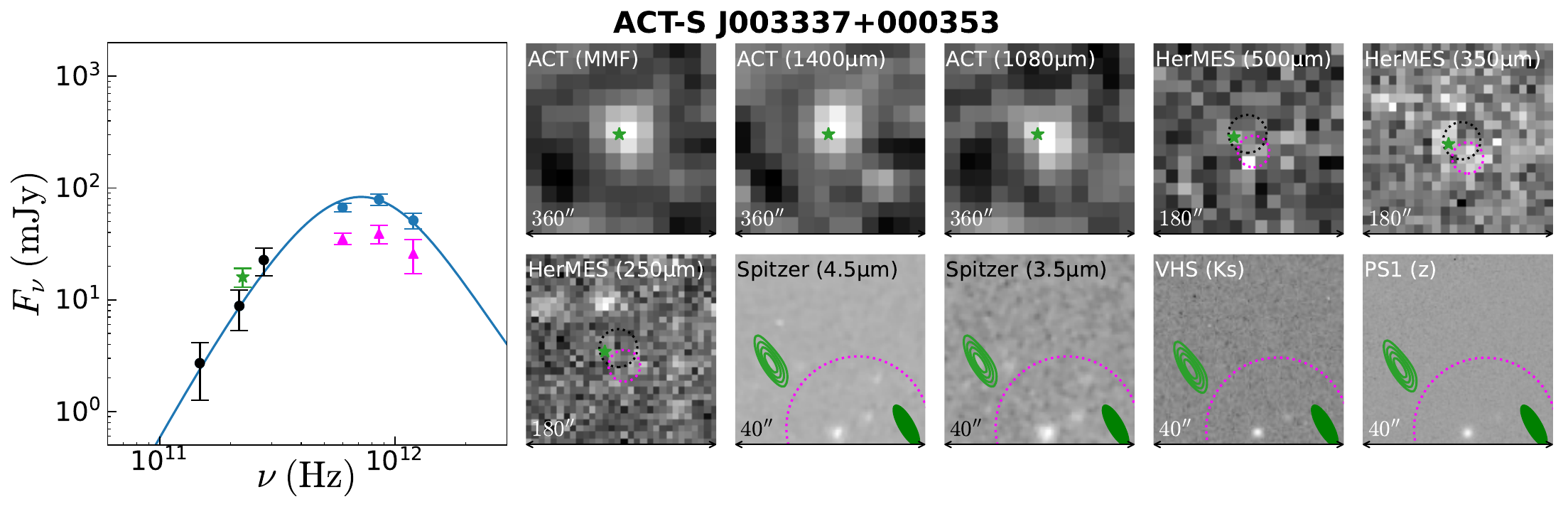}
\includegraphics[width=\columnwidth]{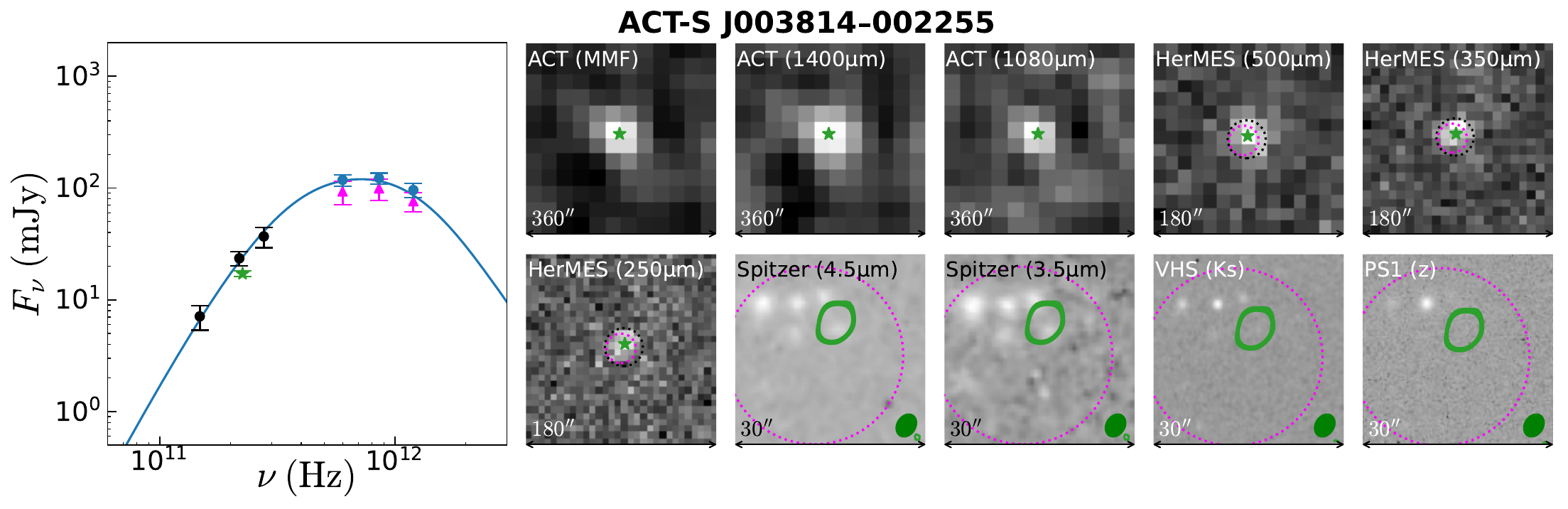}
\includegraphics[width=\columnwidth]{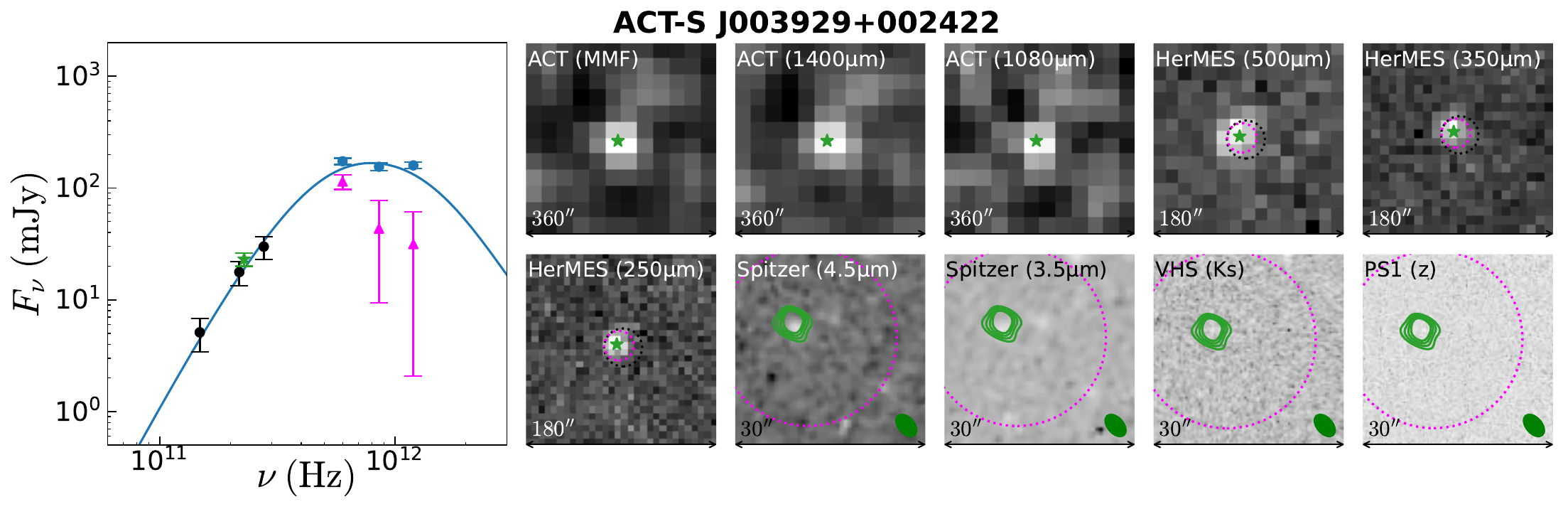}
\caption{Continued}
\label{fig:thumbnailb}
\end{figure*}

\begin{figure*}
\ContinuedFloat
\includegraphics[width=\columnwidth]{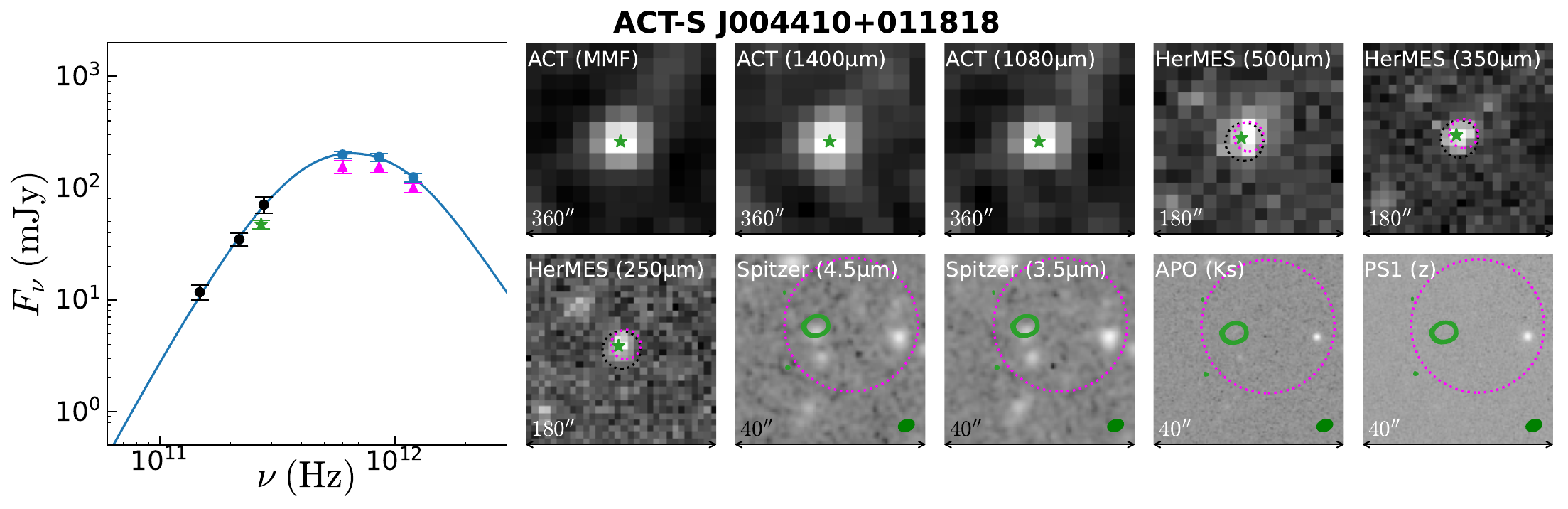}
\includegraphics[width=\columnwidth]{J004532-000127.pdf}
\includegraphics[width=\columnwidth]{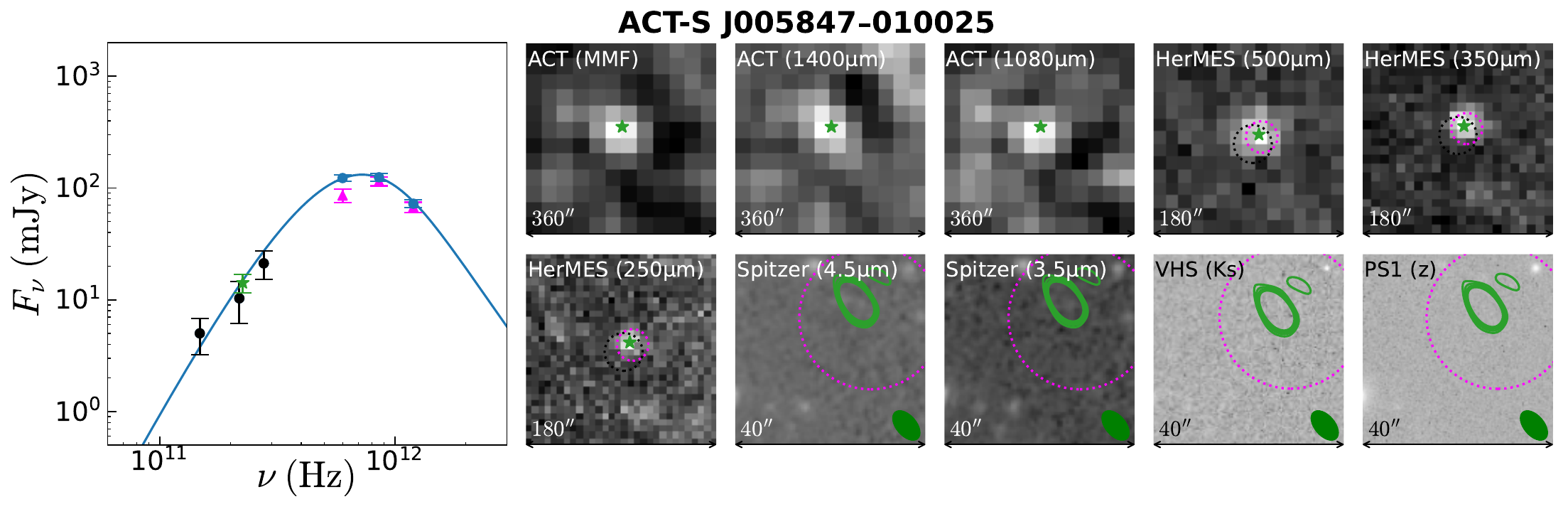}
\includegraphics[width=\columnwidth]{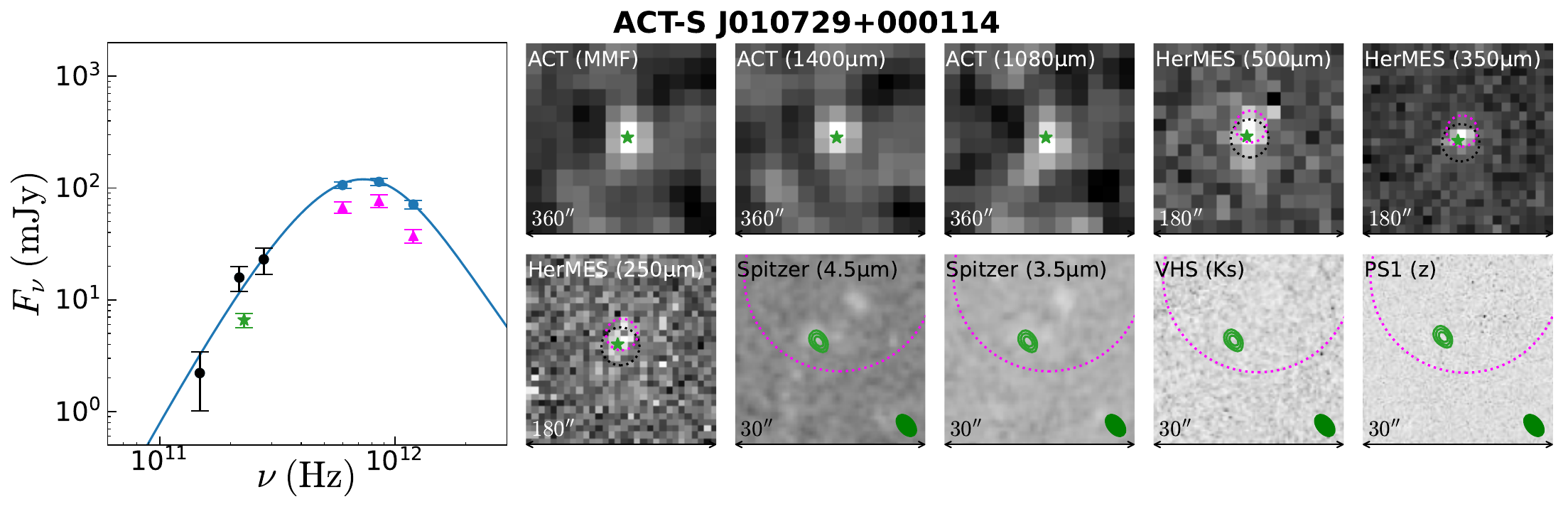}
\caption{Continued}
\label{fig:thumbnailc}
\end{figure*}

\begin{figure*}
\ContinuedFloat
\includegraphics[width=\columnwidth]{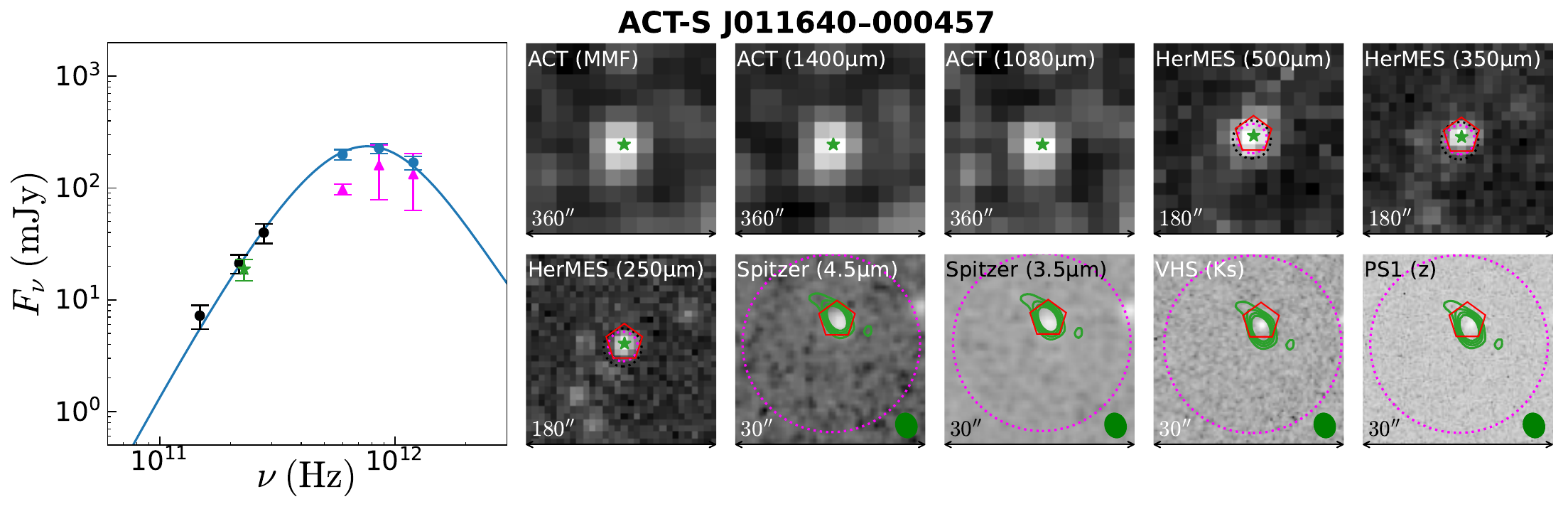}
\includegraphics[width=\columnwidth]{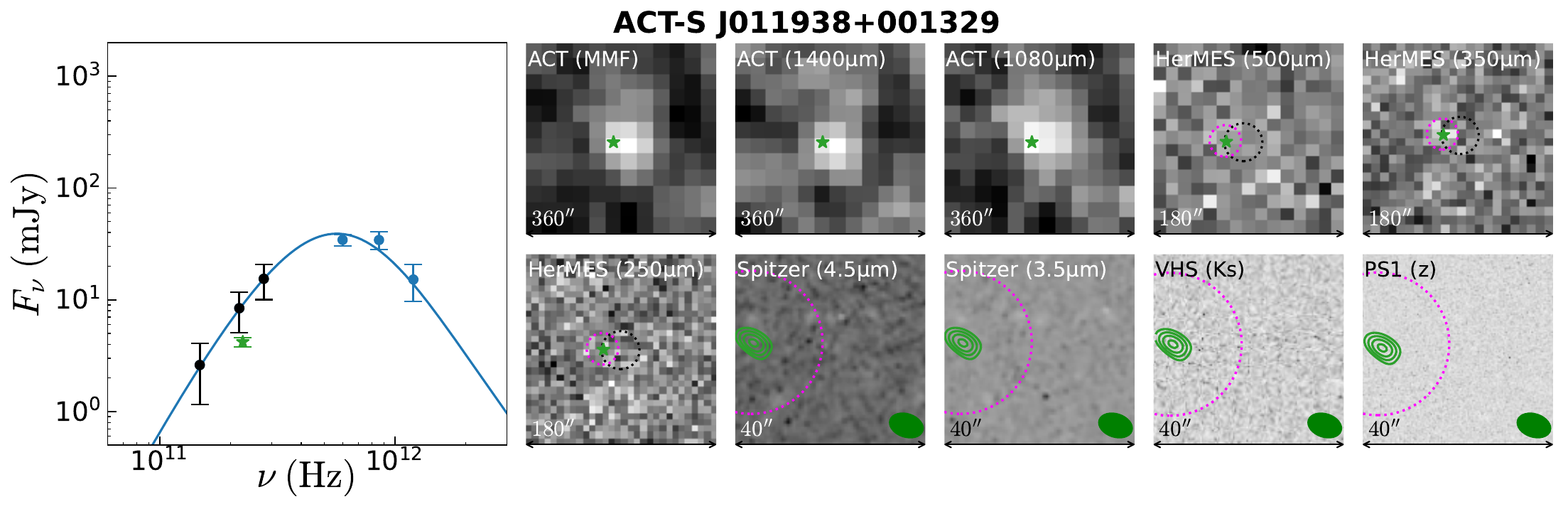}
\includegraphics[width=\columnwidth]{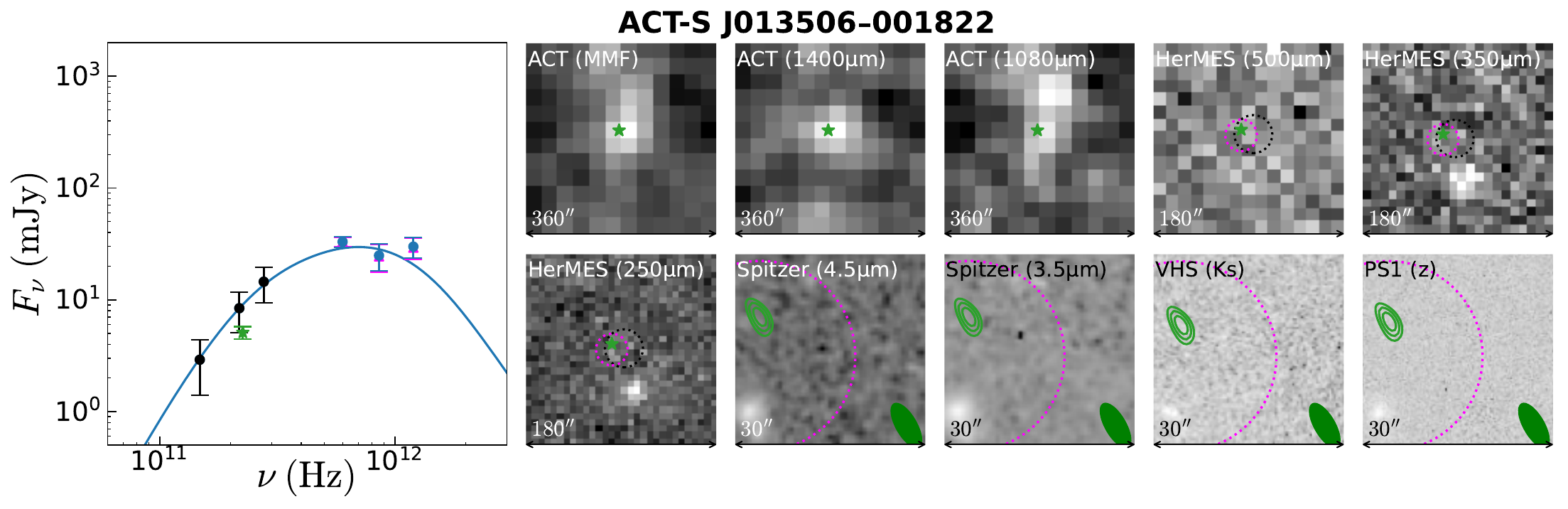}
\includegraphics[width=\columnwidth]{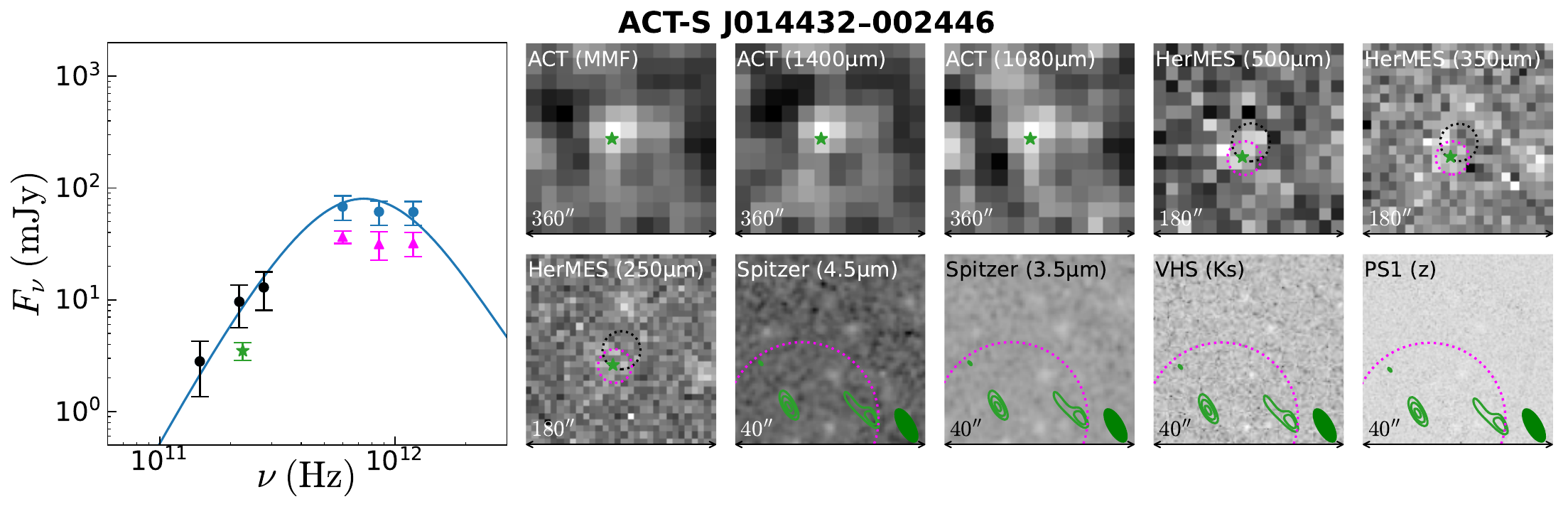}
\caption{Continued}
\label{fig:thumbnailc}
\end{figure*}

\begin{figure*}
\ContinuedFloat
\includegraphics[width=\columnwidth]{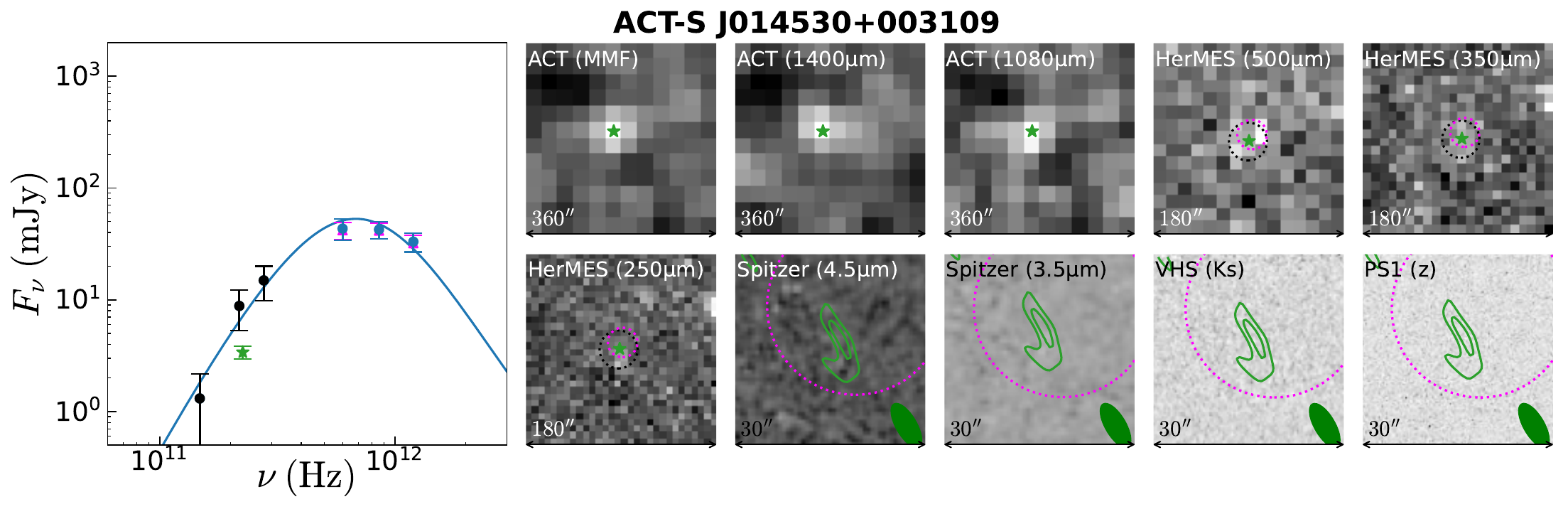}
\includegraphics[width=\columnwidth]{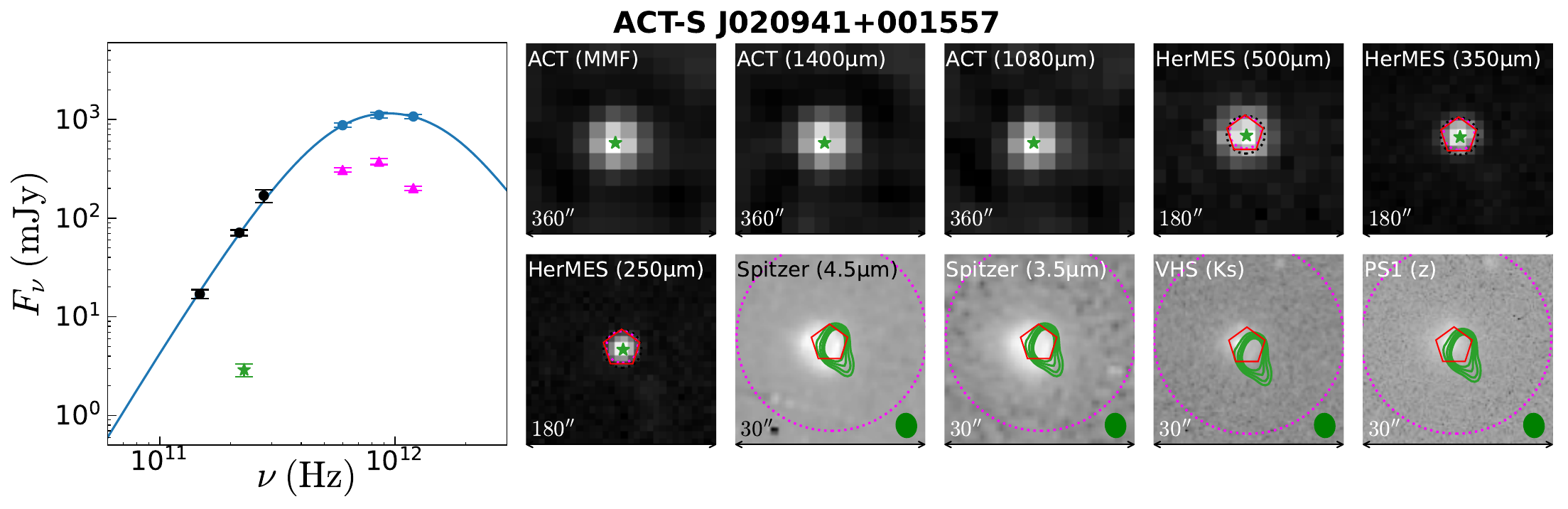}
\includegraphics[width=\columnwidth]{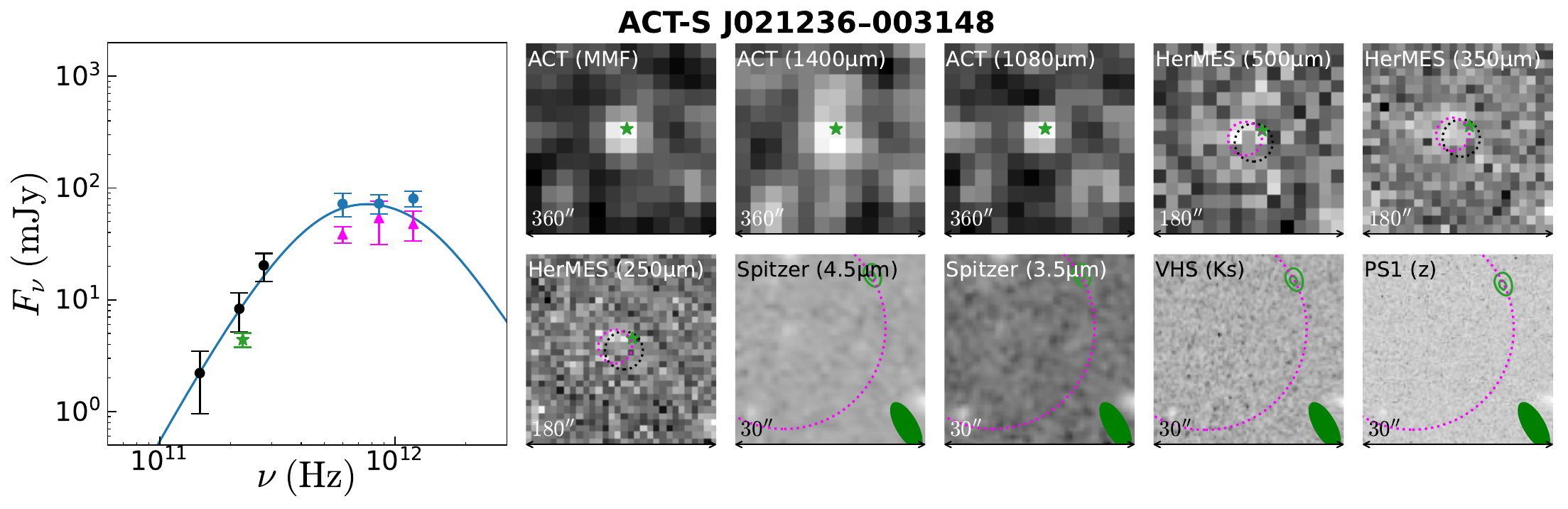}
\includegraphics[width=\columnwidth]{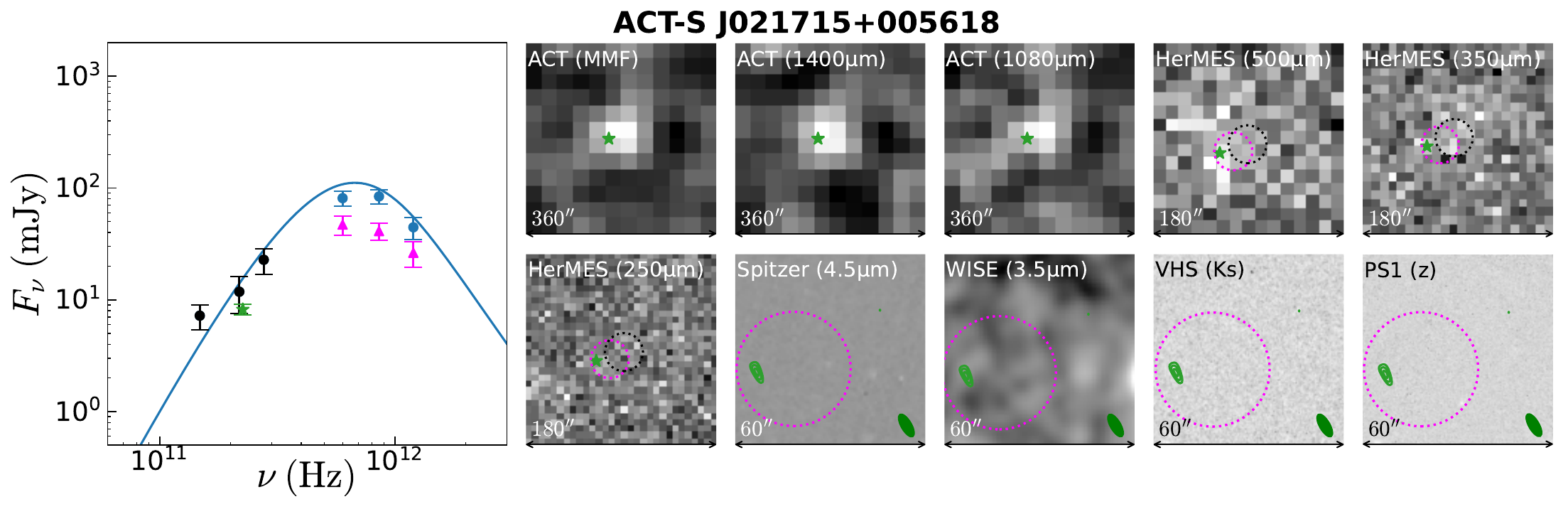}
\caption{Continued}
\label{fig:thumbnailc}
\end{figure*}

\end{document}